\documentclass[]{spie}               

\usepackage{amsmath,amsfonts,amssymb}
\usepackage{graphicx}
\usepackage[colorlinks=true, allcolors=blue]{hyperref}

\title{Cosmological surveys with multi-object spectrographs}

\author[]{Matthew Colless}
\affil[]{Research School of Astronomy and Astrophysics, Australian
  National University,\newline Canberra, Australia}

\authorinfo{Author email: matthew.colless@anu.edu.au}

\begin{document} 
\maketitle

\begin{abstract}
  Multi-object spectroscopy has been a key technique contributing to the
  current era of `precision cosmology'. From the first exploratory
  surveys of the large-scale structure and evolution of the universe to
  the current generation of superbly detailed maps spanning a wide range
  of redshifts, multi-object spectroscopy has been a fundamentally
  important tool for mapping the rich structure of the cosmic web and
  extracting cosmological information of increasing variety and
  precision. This will continue to be true for the foreseeable future,
  as we seek to map the evolving geometry and structure of the universe
  over the full extent of cosmic history in order to obtain the most
  precise and comprehensive measurements of cosmological parameters.
  Here I briefly summarize the contributions that multi-object
  spectroscopy has made to cosmology so far, then review the major
  surveys and instruments currently in play and their prospects for
  pushing back the cosmological frontier. Finally, I examine some of the
  next generation of instruments and surveys to explore how the field
  will develop in coming years, with a particular focus on specialised
  multi-object spectrographs for cosmology and the capabilities of
  multi-object spectrographs on the new generation of extremely large
  telescopes.
\end{abstract}


\section{INTRODUCTION}
\label{sec:intro}

Multi-object spectroscopy (MOS) has proved to be a richly fertile
technique for probing the large-scale structure of the universe, from
which it has proved possible to measure a wide range of cosmological
parameters with increasing precision. For cosmological purposes, the key
aspects of MOS are multiplex (the number of sources that can be
simultaneously observed) and field of view (the area of sky that can be
accessed at one time), since the primary requirements for cosmological
surveys are the size and volume of the sample. Usually (though not
invariably), the quantity being determined for each source is simply the
redshift---the redward shift of spectral features resulting from the
expansion of the universe during the time between emission and
observation. This is (again, usually) a relatively undemanding
measurement to make, requiring neither high signal-to-noise in the data
nor sophistication in the analysis. Hence cosmological surveys typically
reduce galaxy spectra to their redshifts and focus all their efforts on
covering as large a sample as possible over as large a volume as
possible.

The two key milestones in the use of MOS surveys for cosmology were,
first, the step up to surveys that covered a
statistically-representative volume of the relatively local universe
and, second, the further step up to surveys that covered
statistically-representative volumes over cosmologically significant
ranges in redshift. The first step ushered in the era of `precision
cosmology' for MOS surveys, while the second step made MOS surveys a key
tool for probing the nature of dark energy.

\begin{figure}
  \begin{center}
    \includegraphics[width=0.48\textwidth]{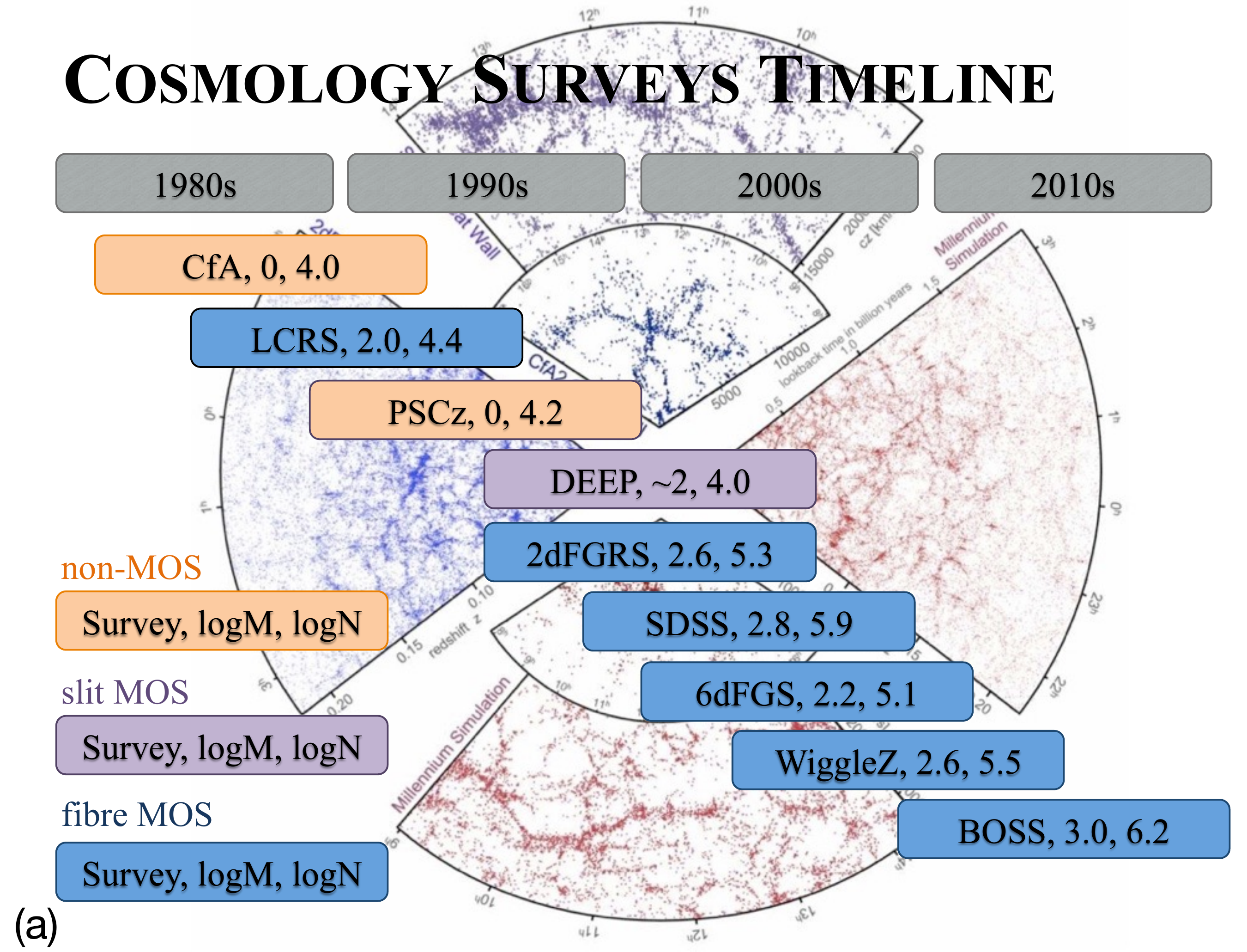}
    \includegraphics[width=0.48\textwidth]{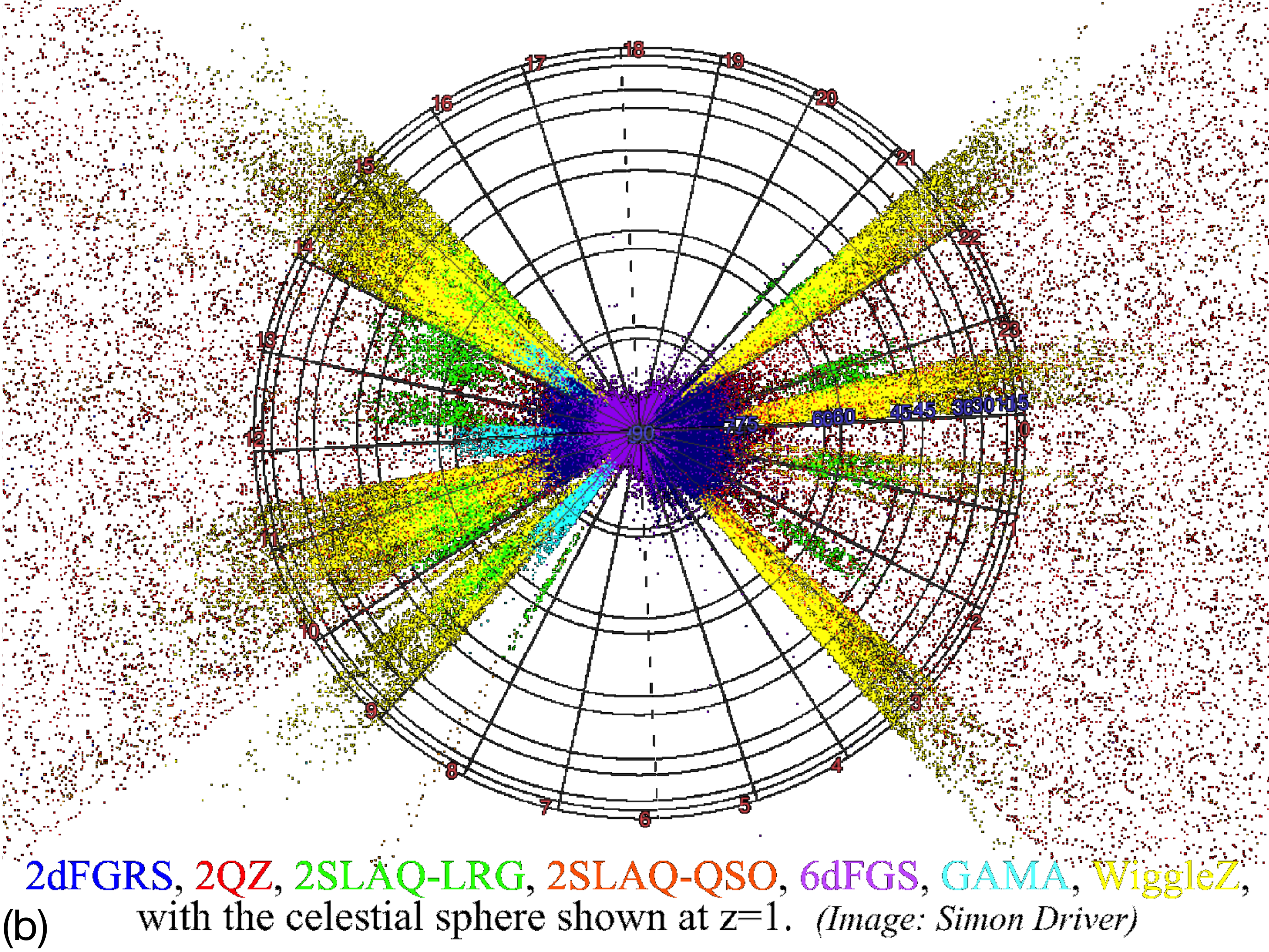}
  \end{center}
  \caption{\label{fig:timelinemap}(a)~A schematic history of
    cosmological redshift surveys, showing the approximatw order and
    dates of key surveys. The labels give the survey name and the log of
    the multiplex and sample size ($\log M$ and $\log N$ resp.); the
    colour indicates whether the survey was non-MOS (orange), slit-MOS
    (purple) or fibre-MOS (blue). (b)~The redshift maps for the main MOS
    surveys carried out with AAO facilities.}
\end{figure}

In the following sections I briefly review the past, present, and future
of cosmological MOS surveys. An overview of the key cosmology surveys is
provided in Figure~\ref{fig:timelinemap}, which gives a timeline of
significant surveys showing the increasing multiplex and survey size
resulting from the ongoing development of multi-object spectrograph
systems (especially fibre MOS), and a map of the large-scale structure
in the universe based on the various redshift surveys carried out with
Australian Astronomical Observatory (AAO) facilities over the last 20
years.

\section{THE CLASSICAL PERIOD}
\label{sec:classical}

Due to the modest field of view and multiplex of the early MOS
instruments on (mostly) 4-metre class telescopes, the initial
3applications of MOS surveys tended to be studies of relatively small
volumes (often clusters of galaxies\cite{Colless1987}) or `pencil-beam'
surveys over a limited range of redshifts (usually looking at galaxy
evolution\cite{Broadhurst1988, Colless1990, Ellis1996}, even though
$z \ll 1$).

The pioneering surveys by Geller, Huchra and
collaborators\cite{deLapparent1986, Geller1989, deLapparent1991}---which
were done {\it without} the use of MOS instruments---demonstrated that
there was a rich structure in the large-scale distribution of galaxies.
As the theory of cosmological density perturbations\cite{Kodama1984,
  Mukhanov1992} was developed, it was recognized that this rich
structure encoded key cosmological parameters such as the overall
density of the universe, and moreover could be used to distinguish the
relative contributions of ordinary `baryonic' matter, non-relativistic
`cold' dark matter, and relativistic `hot' dark matter. This
potentiality motivated the first MOS surveys with the explicit goal of
measuring cosmological cosmological parameters, in contrast to previous
surveys that either sought to explore the structure of the large-scale
galaxy distribution or the evolution of the galaxy population.

The most extensive of this first generation of cosmological MOS surveys
were the Las Campanas Redshift Survey\cite{Shectman1996, Landy1996,
  Lin1996, Tucker1997, Landy1998} (LCRS), and the IRAS Point Source
Catalogue Redshift Survey\cite{Canavezes1998, Tadros1999, Schmoldt1999,
  Rowan-Robinson2000, Saunders2000} (PSCz). The latter was a sparse
all-sky survey that did not use MOS, but the LCRS was an important
forerunner to all subsequent cosmological MOS surveys. It surveyed more
than 26,000 galaxies over an area of 700\,deg$^2$ and reached a median
depth of $z \approx 0.1$. The LCRS took 6~years on the Las Campanas
2.5-metre du Pont telescope, which had a MOS system with a multiplex of
112 and a field of view 2.1\,deg in diameter. It was an eye-opener for
observational cosmologists, demonstrating the potential of MOS for
probing large-scale structure and measuring the galaxy luminosity
function, the correlation function and power spectrum (in 2D and 3D),
the pairwise velocity distribution, and producing a catalogue of groups
and clusters. However it also illuminated the size of the
challenge---despite being the largest redshift survey up to that time,
it was nonetheless still an order of magnitude smaller than a survey of
a statistically-representative volume of the universe needed to be.

\section{THE ENLIGHTENMENT}
\label{sec:enlightenment}

Two groups took on the challenge of constructing the first truly
cosmological survey of the nearby universe. 

\begin{figure}
  \begin{center}
    \includegraphics[width=0.58\textwidth]{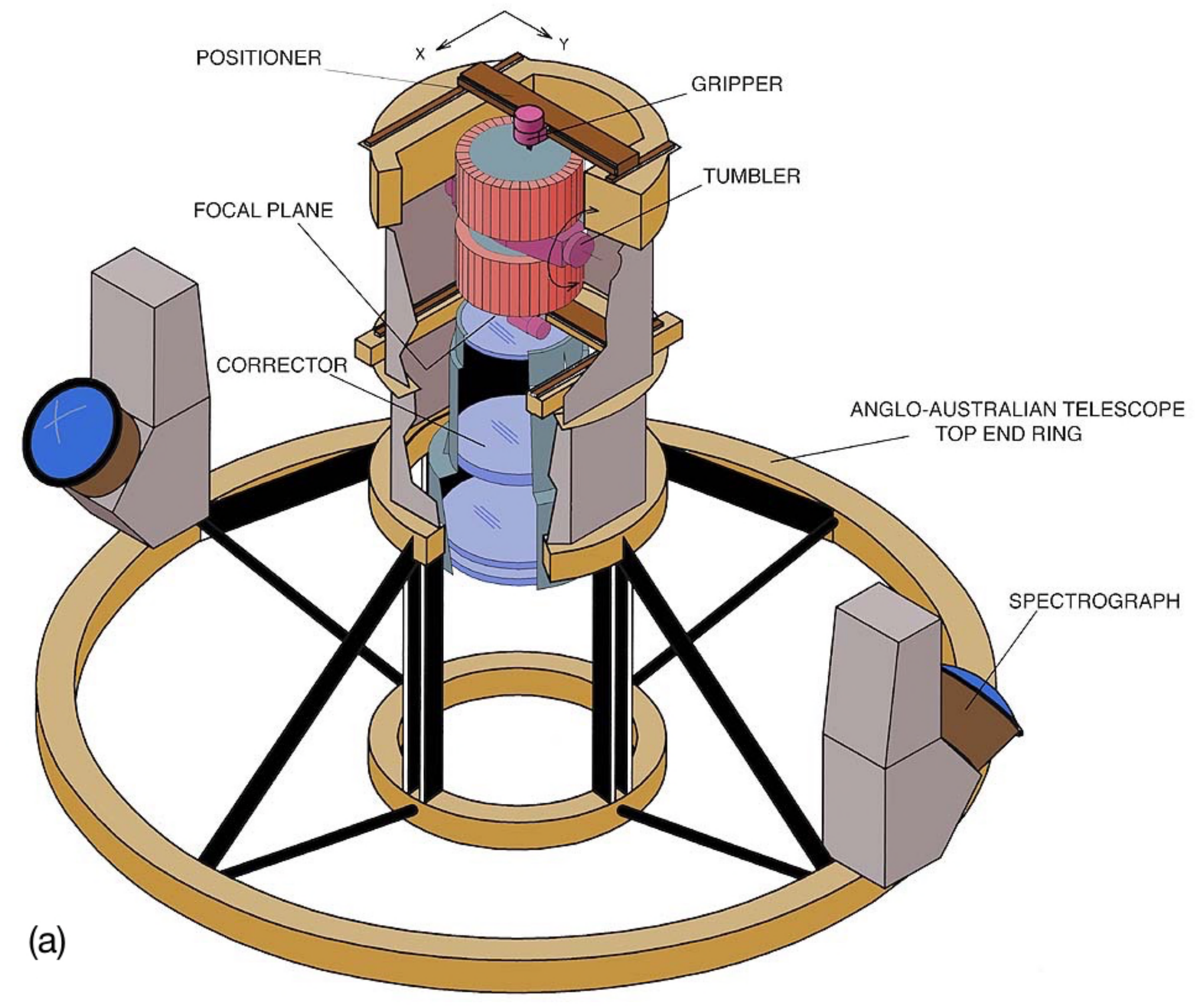}
    \includegraphics[width=0.40\textwidth]{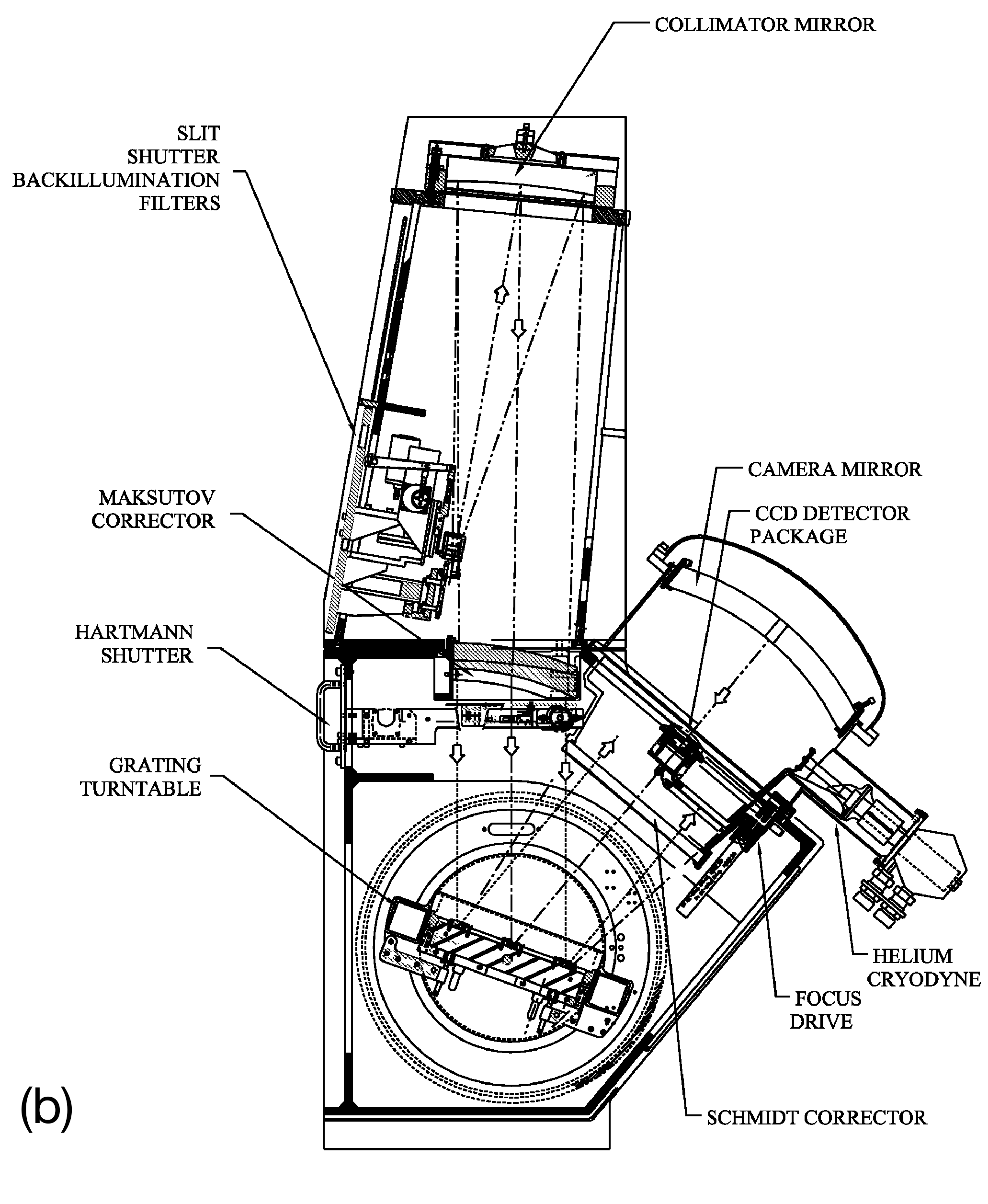}
  \end{center}
  \caption{\label{fig:2dFinst}(a)~Schematic of the 2dF topend ring for
    the 3.9-metre Anglo-Australian Telescope (AAT), showing the location
    of the two spectrographs, and a cutaway view of the corrector and
    positioner system. (b)~A section view of one of the 2dF
    spectrographs showing the main subsystems.}
\end{figure}

One was an British and Australian team that built a revolutionary new
MOS, the 2-degree Field multi-fibre spectrograph (2dF) for the 4-metre
Anglo-Australian Telescope\cite{Cannon1989}. As its name implies, 2dF
had a 2-degree diameter field of view (the largest on any on a
4-metre-class telescope), a multiplex of 400, and used a robotic fibre
positioner that facilitated rapid automatic
reconfiguration\cite{Taylor1990, Gray1990}. The 2dF fibre positioning
system was considered technically challenging and risky at the time it
was being constructed (the early 1990s), but it proved effective and
reliable in practice\cite{Lewis2002}. 2dF's wide field of view was
achieved with a corrector lens incorporating an atmospheric dispersion
compensator, which was an essential innovation in a system aiming to
achieve wide spectral coverage with small aperture fibres. The robot
positioner placed fibres sequentially at the rate of one every 6\,s with
a precision of 0.3\,arcsec (corresponding to 20\,$\mu$m).
Figure~\ref{fig:2dFinst}a shows the 2dF topend ring, with the corrector,
positioner and the two spectrographs. Because it took about an hour to
reconfigure a complete set of fibres, 2dF had a double-buffering system
with two field plates each having 400 fibres. While the fibres on one
field plate were being reconfigured by the robot, the second field plate
was being observed; at the end of an observation the two field plates
were tumbled into the other position and the process repeated. The
2.1\,arcsec (140\,$\mu$m) diameter fibres fed a pair of dual-channel
spectrographs that offered spectral resolving powers
($R$=$\lambda$/FWHM) between $R$=500 and $R$=2000 covering wavelength
ranges of 440\,nm and 110\,nm respectively. The overall throughput of
the entire 2dF system peaked at about 5\% around 600\,nm. The layout of
the 2dF spectrographs is shown in Figure~\ref{fig:2dFinst}b.

The 2dF MOS system was designed from the outset to enable a massive
galaxy redshift survey to test the cosmological model and measure its
key parameters. Over a period of 5 years from 1997 to 2002, the 2dF
Galaxy Redshift Survey (2dFGRS) measured 221,000 redshifts over
$\sim$1500\,deg$^2$ with a median depth of $z \sim 0.11$ (corresponding
to a volume of $\sim$0.12\,Gpc$^3$), making it the first cosmological
redshift survey to capture a statistically-representative sample of the
universe\cite{Colless2001, Colless2003}.

The main cosmological results from 2dFGRS related to the nature of the
large-scale structure, the overall density of the universe, and the
nature of its massive constituents\cite{Peacock2001, Percival2001,
  Efstathiou2002, Elgaroy2002, Cole2005, Colless2011}. The survey
precisely determined the statistical properties of the large-scale
structure of the galaxy distribution (via the galaxy power spectrum or,
equivalently, the galaxy correlation function) over size scales from
about 1\,Mpc to about 300\,Mpc. The properties of the galaxy
distribution confirmed the generally accepted paradigm that the
large-scale structure grows by gravitational instability in a way that
is qualitatively and quantitatively consistent with the standard model
of gravitational amplification of quantum fluctuations emerging from the
Big Bang. From the power spectrum and redshift-space distortions, 2dFGRS
obtained an estimate for the total density of all types of matter in the
universe of $\Omega_M = 0.23 ± 0.02$; the uncertainty of less than 10\%
on this figure was one of major steps towards ‘precision cosmology’ from
redshift surveys. Moreover, 2dFGRS was able to show that the fraction of
the total matter density in baryons is 18\%, consistent with a baryon
density of $\Omega_B = 0.04$, as found from the cosmic microwave
background anisotropies and Big Bang nucleosynthesis models. On the
other hand, relativistic matter such as neutrinos makes up less than
13\% of the overall matter density, implying an upper limit on the total
mass of the three neutrino species of 0.7\,eV.

\begin{figure}
  \begin{center}
    \includegraphics[width=0.48\textwidth]{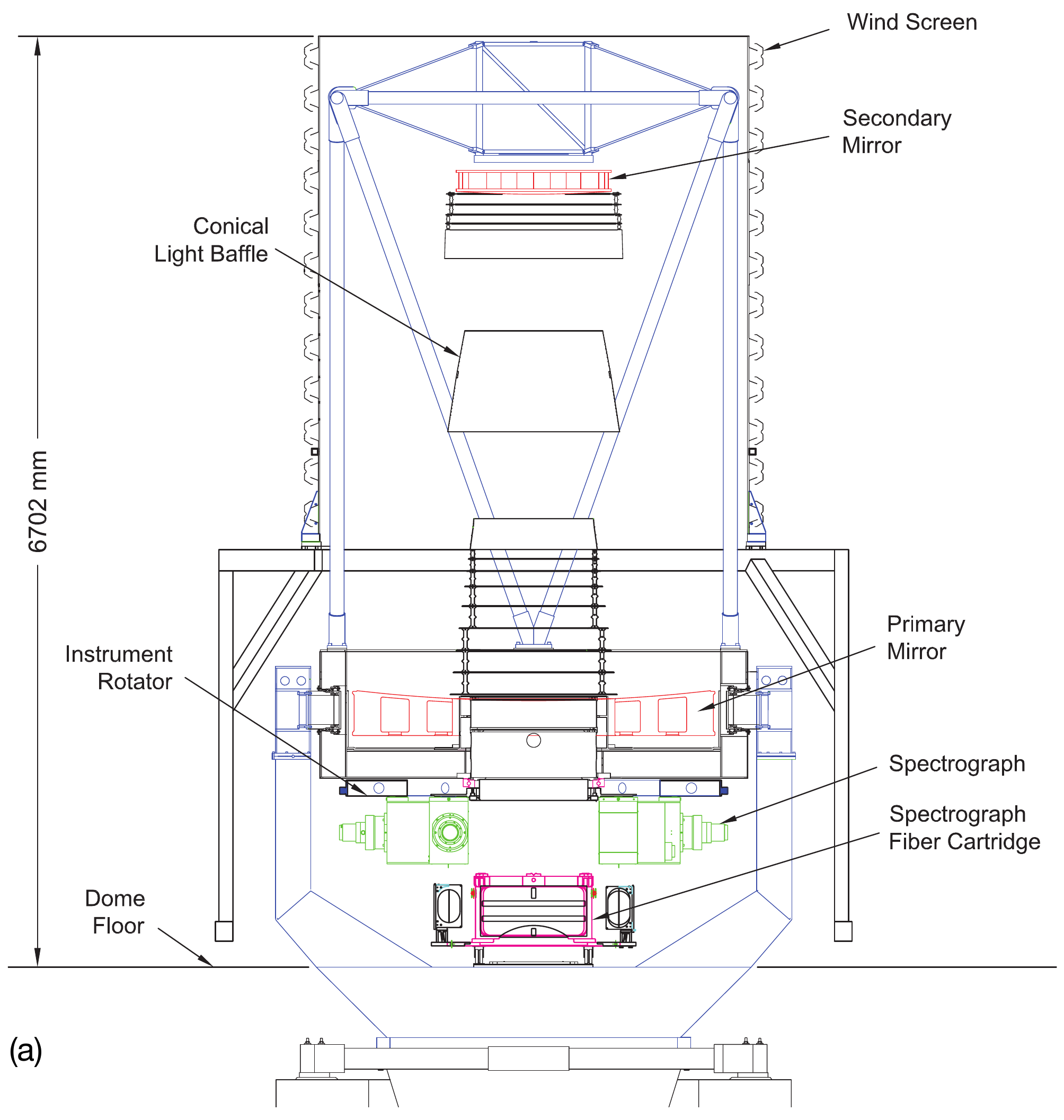}
    \includegraphics[width=0.48\textwidth]{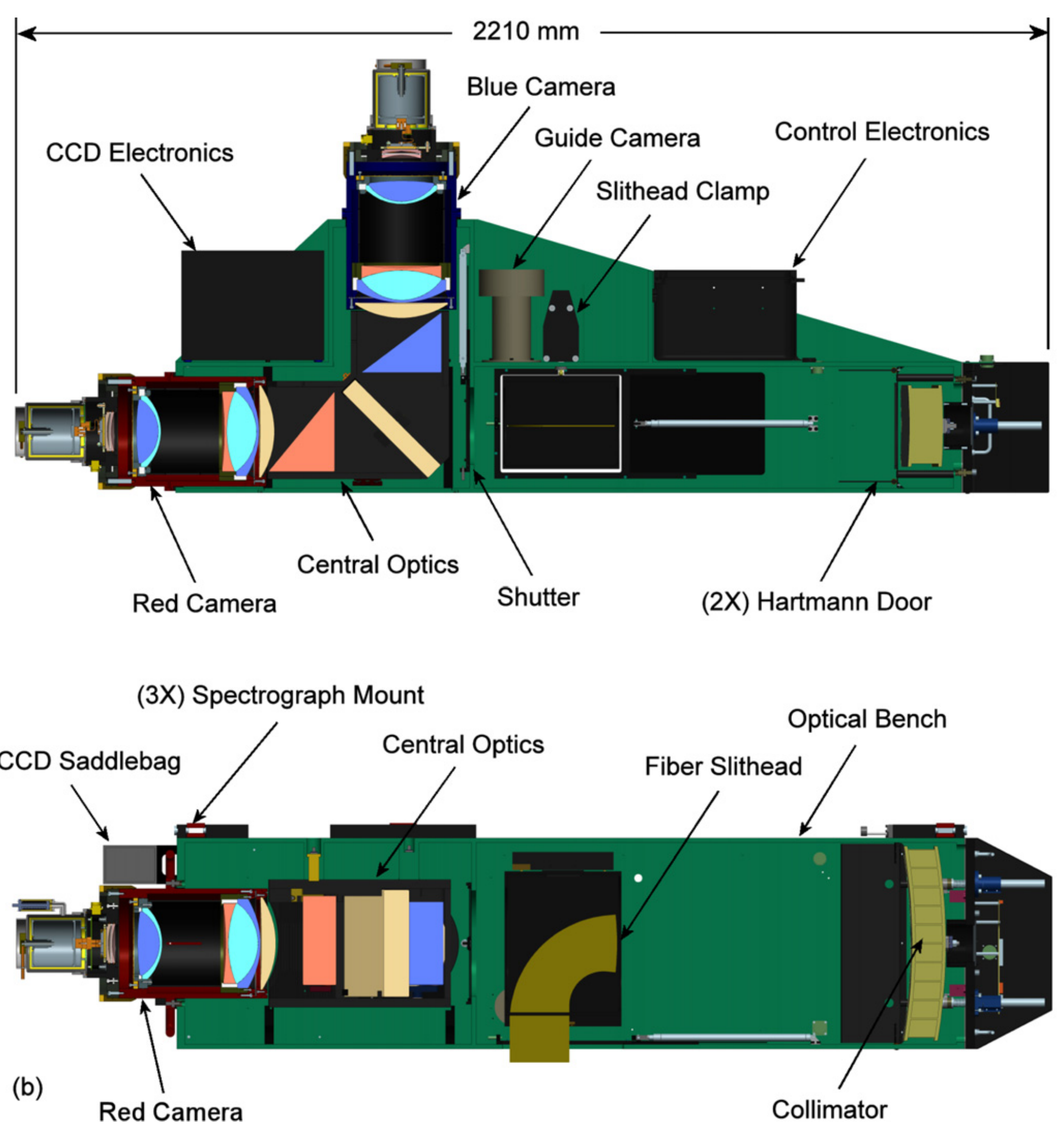}
  \end{center}
  \caption{\label{fig:SDSSinst}(a)~Schematic of the Apache Point
    Observatory 2.5-metre telescope showing the location of the
    spectrographs\cite{Smee2013}, with a fiber cartridge in the
    retracted position. The twin spectrographs mount to the back of the
    Cassegrain instrument rotator adjacent to the focal plane with
    sufficient separation to allow installation and removal of the
    imaging camera and fiber cartridges. (b)~Two views of the SDSS
    spectrographs\cite{Smee2013} and the main opto-mechanical
    subassemblies: the fiber slithead, the collimator, the central
    optics, and the red and blue channel cameras.}
\end{figure}

Contemporaneously with the 2dFGRS, an even more ambitious project, the
Sloan Digital Sky Survey (SDSS), was being carried forward by a largely
US-based team\cite{York2000, Stoughton2002}. Whereas 2dFGRS relied on
photographic sky surveys for its input target catalogue, the SDSS
project paired a CCD imaging survey with a MOS spectral survey. SDSS
used a purpose-built 2.5-metre telescope with a 3\,deg diameter field of
view at Apache Point Observatory, and its MOS instrument\cite{Smee2013}
had 640 fibres, each 3\,arcsec (180\,$\mu$m) in diameter, that could be
positioned over the 7\,deg$^2$ field (see Figure~\ref{fig:SDSSinst}a).
Unlike 2dF's robotic system, the SDSS fibre system was a plug-plate
design requiring manual positioning of the fibres. The twin SDSS
spectrographs (see Figure~\ref{fig:SDSSinst}b) were each fed by 320
fibres, and utilized a simple optical layout with reflective
collimators, gratings, all-refractive cameras, and state-of-the-art CCD
detectors to record the spectra from these fibres simultaneously in two
channels over the wavelength range from 390\,nm to 910\,nm at a
resolving power $R \approx 2000$. The overall efficiency of the
spectrographs peaked at about 17\% in the blue channel and 22\% in the
red channel.

The original SDSS survey\cite{Abazajian2009} (comprising SDSS-I and
SDSS-II) ran for 9~years from 2000 to 2008. It imaged an area of
11,663\,deg$^2$ and obtained galaxy spectra over 8032\,deg$^2$,
measuring redshifts for 930,000 galaxies with a median redshift of
$z \approx 0.1$ (corresponding to $\sim$0.5\,Gpc$^3$). The cosmological
results from SDSS\cite{Zehavi2002, Tegmark2004, Eisenstein2005,
  Percival2007} covered similar ground to those of 2dFGRS, but
ultimately achieved higher precision due to both the larger size of the
SDSS sample and the better quality of the CCD imaging and photometry.

With the 2dFGRS and SDSS surveys, MOS spectroscopy had made observations
of large-scale structure a tool for `precision cosmology'---though that
initial level of `precision' seems rather imprecise by today's
standards. One of the key outcomes from these surveys was the detection
of the baryon acoustic oscillation (BAO) signature in the galaxy
distribution\cite{Cole2005, Eisenstein2005, Percival2007}. The BAO
`standard ruler' has been a key tool in subsequent cosmological surveys
seeking to probe the nature of dark energy, using the evolution of the
expansion rate and geometry of the universe as means of determining the
dark energy equation of state.

\section{THE MODERN ERA}
\label{sec:modern}

Following the outstanding success of the 2dFGRS and SDSS-I/II surveys
there has been a procession of follow-on cosmological MOS surveys
addressing a range of different issues. Table~\ref{tab:surveylist}
lists the main recent MOS surveys (both cosmological and other) and
the telescope/instrument combinations with which they have been
carried out. Notable amongst the various cosmological MOS surveys in
this list are the subsequent SDSS surveys (including BOSS and eBOSS),
the WiggleZ survey on the Anglo-Australian Telescope, and the 6dFGS
survey on the UK Schmidt Telescope.

\begin{table}
\caption{\label{tab:surveylist}A list of the telescope:instrument
  combinations used to perform various MOS surveys (cosmological and
  other).}
  \begin{center}
    \includegraphics[width=0.67\textwidth]{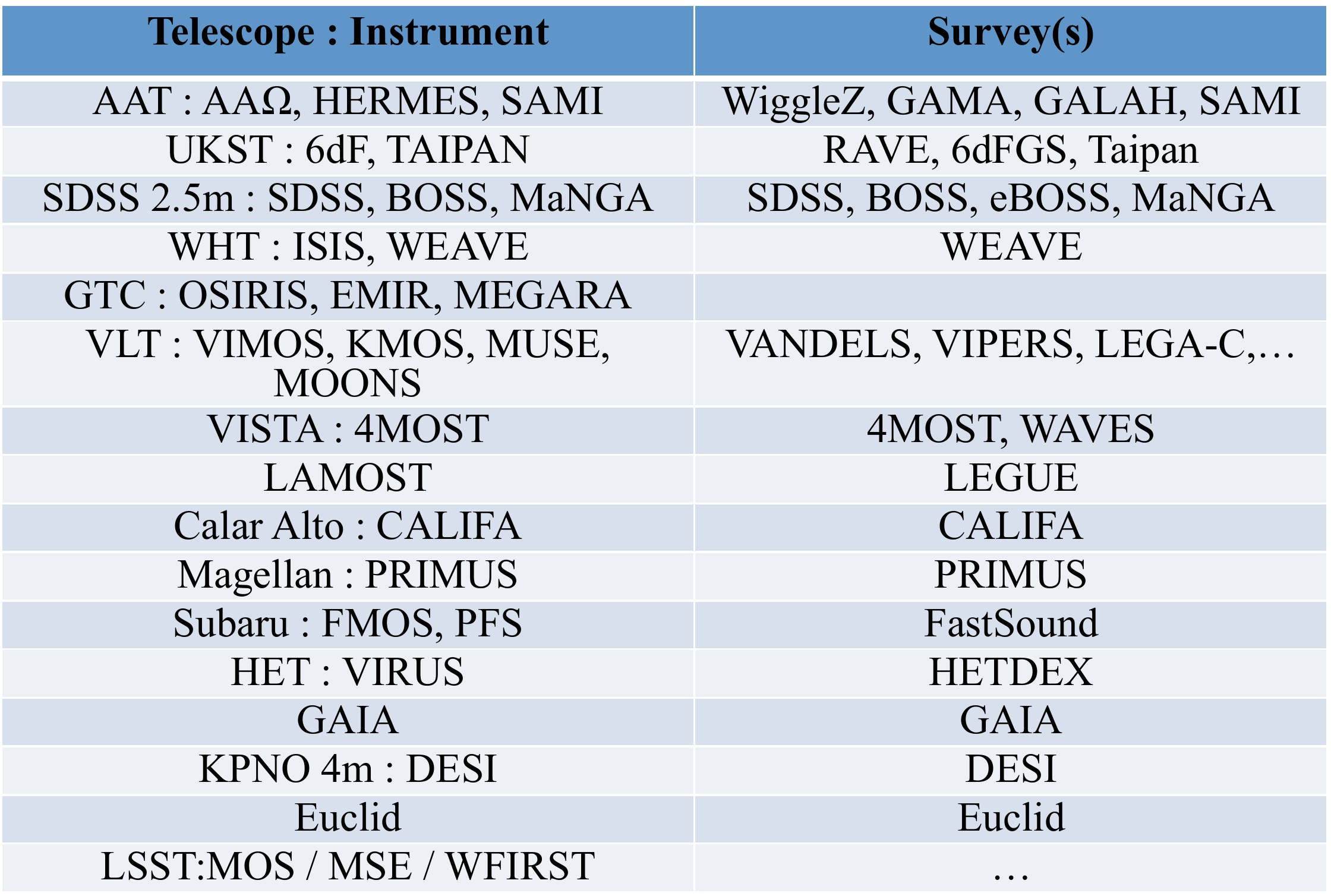}
  \end{center}
\end{table}

\subsection{6dFGS}
\label{sec:6dFGS}

The 6dF Galaxy Survey (6dFGS) is a cosmological survey of the relatively
local ($z<0.1$) universe\cite{Wakamatsu2003, Jones2004, Colless2005,
  Colless2006}, differing from others in combining a redshift survey
with a peculiar velocity survey. The redshift survey\cite{Jones2004,
  Jones2009} was relatively standard, except that it covered whole
southern hemisphere apart more than 10\,deg from the Galactic Plane (at
total area of 17,000\,deg) and was restricted to the brightest 125,000
galaxies in the local universe. The peculiar velocity
survey\cite{Magoulas2012, Campbell2014, Springob2014} used Fundamental
Plane distances for early-type galaxies, in combination with their
redshifts, to measure the peculiar velocities (i.e.\ non-Hubble-flow
motions) for about 8000 galaxies at distances
$cz < 16,000$\,km\,s$^{-1}$. Peculiar velocity surveys, by adding direct
measurements of the motions of galaxies due to the effective
gravitational force of the surrounding matter, provide additional,
complementary information to redshift surveys regarding both the mass
distribution and the nature of gravity. However, available methods for
measuring galaxies' distances independent of their redshifts are not
precise (e.g.\ the Fundamental Plane provides distances with typical
errors of about 20\%), and as a result they can only measure peculiar
velocities effectively at relatively small distances (i.e.\ relatively
low redshifts).

\begin{figure}
  \begin{center}
    \includegraphics[width=\textwidth]{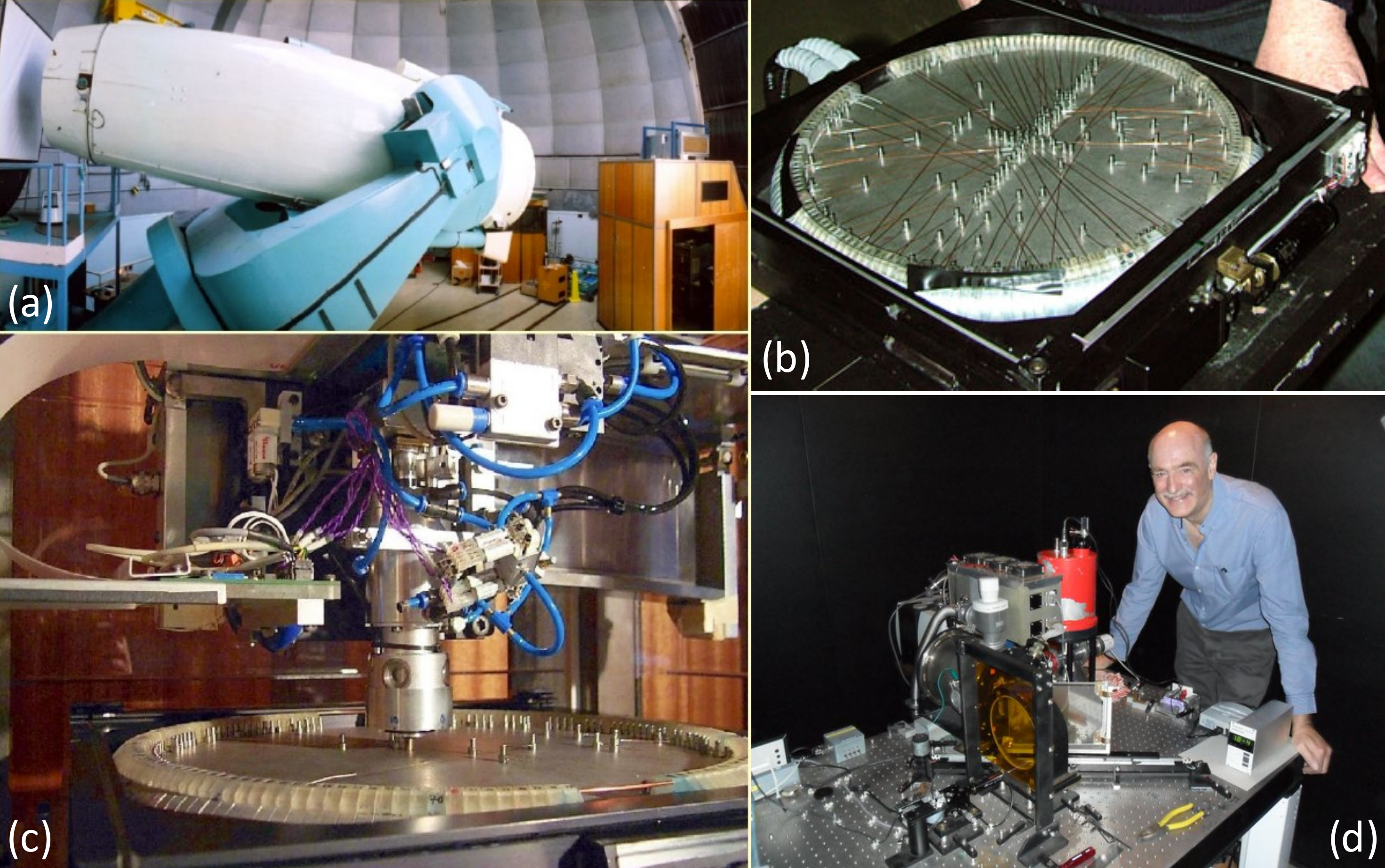}
  \end{center}
  \caption{\label{fig:6dFinst}Images of the 6dF instrument. (a)~The UK
    Schmidt Telescope (UKST) at Siding Spring Observatory. (b)~The 6dF
    field plate holder, showing optical fibres and buttons on the convex
    field plate. (c)~The 6dF $r$--$\theta$--$z$ robot. (d)~The 6dF
    spectrograph with its designer, Fred Watson.}
\end{figure}

The 6dFGS observations were carried out over 5~years from 2001 to 2006
using the 6-degree Field (6dF) MOS system on the UK Schmidt Telescope
(UKST)\cite{Parker1998, Watson1998, Parker2000, Watson2000, Jones2004}.
The 6dF system had 150 science fibers, each 6.7\,arcec (100\,$\mu$m) in
diameter, that could be positioned over the 5.7-degree field of the UKST
(see Figure~\ref{fig:6dFinst}a\,\&\,b). This made 6dF the ideal
instrument for full-sky spectroscopic surveys of relatively sparse
($<$50\,deg$^{-2}$), bright (V$<$17) objects. In terms of the A$\Omega$
(telescope aperture $\times$ field of view) figure of merit, 6dF on the
1.2-metre UKST has about 75\% of the survey power of 2dF on the
3.9-metre Anglo-Australian Telescope (AAT), but with the operational
advantage that it was entirely given over to survey observations.
Because of the curved focal plane of the UKST, 6dF used a 3-axis
$r$--$\theta$--$z$ robot (see Figure~\ref{fig:6dFinst}c) with a curved
radial arm to position the individual fibres, which were contained in an
adapted version of the Schmidt photograph plate holder. The robot was
located off the telescope, and the plate holders had to be manually
mounted in the UKST. The relatively large `buttons' that magnetically
held the fibres on the field plate, combined with the small platescale
of the UKST, meant that targets closer than 5.7\,arcmin could not be
observed simultaneously. The fibres fed a floor-mounted, fixed format
spectrograph (see Figure~\ref{fig:6dFinst}d), which used reflection
gratings up to October 2002 and thereafter volume phase holographic
gratings. The limited size of the CCD detector meant that, while all 150
fibres fitted on the detector simultaneously, full spectral coverage
from $\sim$400\,nm to $\sim$800\,nm required two exposures with
different gratings. The peak system efficiency was about 11\% at
wavelengths near the gratings' blaze angles.

The main cosmological results from the 6dFGS relate to {\em local}
(i.e.\ $z\approx0$) measurements of key parameters, which are important
because they require little or no reliance on an assumed cosmological
model to interpret them. Thus the 6dFGS redshift survey yielded a direct
and independent estimate of the local Hubble constant\cite{Beutler2011}
($H_0$)---based on the standard ruler provided by the baryonic acoustic
oscillations (BAO)---that agrees well with the value obtained by the
Planck satellite extrapolating from the cosmic microwave background
(CMB) at $z \approx 1100$ assuming the standard $\Lambda$CDM cosmology.
The 6dFGS similarly provide low-redshfit measurements of the product of
the normalisation of the matter power spectrum ($\sigma_8$) and the
growth rate of large-scale structure\cite{Beutler2012} derived from the
redshift-space distortions in the two-point galaxy correlation function.
This low-redshift constraint was combined with the very high-redshift
CMB constraint to confirm that the growth of structure over the history
of the universe is consistent with the model for gravity provided by
General Relativity. By contrast, the peculiar velocity measurements have
no high-redshift counterparts, but are complementary to the redshift
survey. After careful calibration of the Fundamental
Plane\cite{Magoulas2012}, the 6dFGS was able to measure peculiar
velocities for nearly 8000 early-type galaxies and map the large-scale
velocity field\cite{Springob2014} in the southern hemisphere out to
$\sim$16,000~km\,s$^{-1}$. This observed velocity field was used to
determine the rms bulk motions\cite{Scrimgeour2016} on scales of
50--70\,$h^{-1}$\,Mpc and to make a direct comparison with the predicted
velocity field derived from the redshift survey density
field\cite{Magoulas2016}. The 6dFGS peculiar velocities also allowed the
first-ever direct measurement of the velocity power
spectrum\cite{Johnson2014}, which provided another means of measuring
the growth rate of structure and showing that it is scale-independent up
to scales of at least 300\,$h^{-1}$\,Mpc.

\subsection{WiggleZ}
\label{sec:WiggleZ}

\begin{figure}
  \begin{center}
    \includegraphics[width=0.42\textwidth]{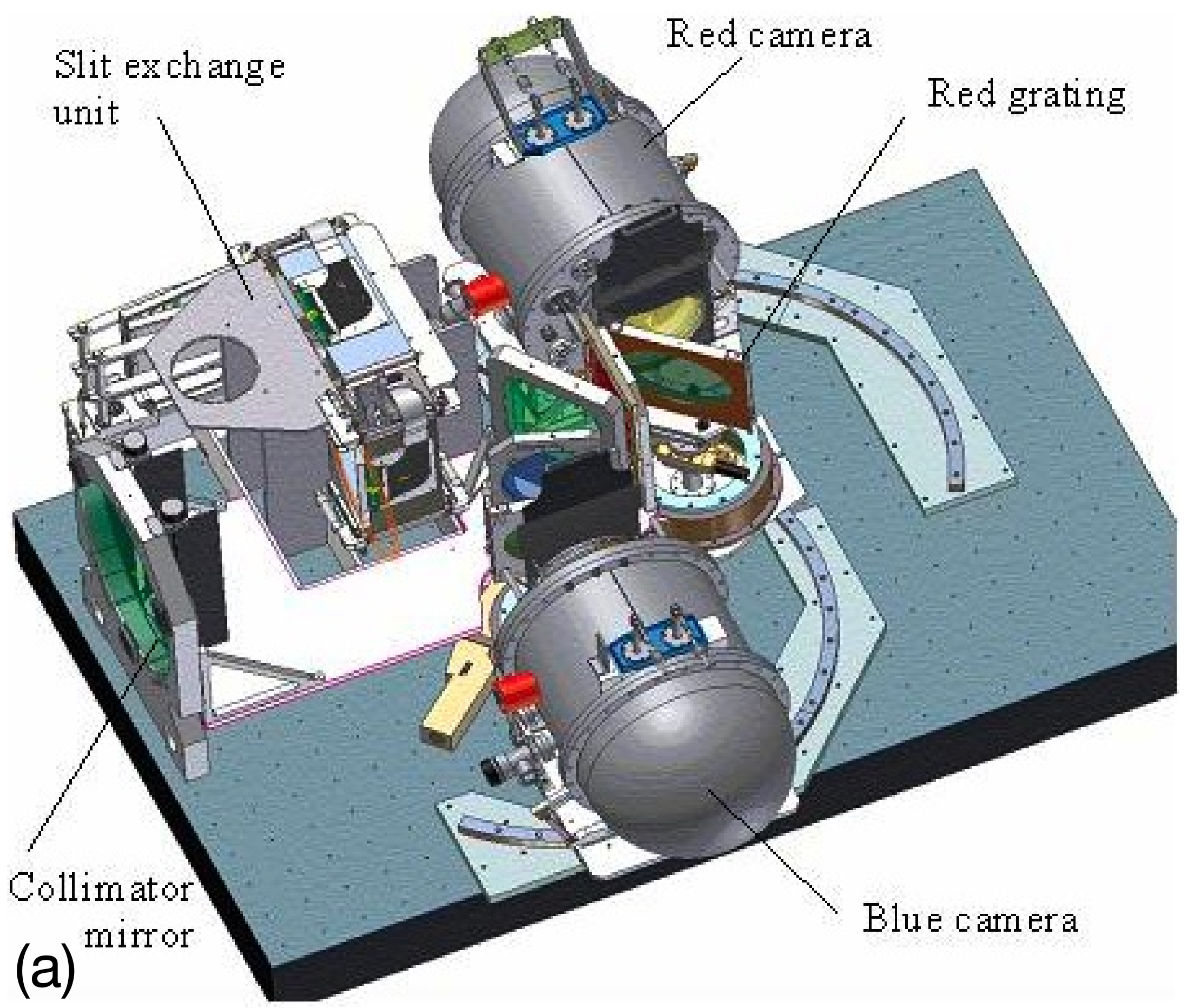}
    \includegraphics[width=0.54\textwidth]{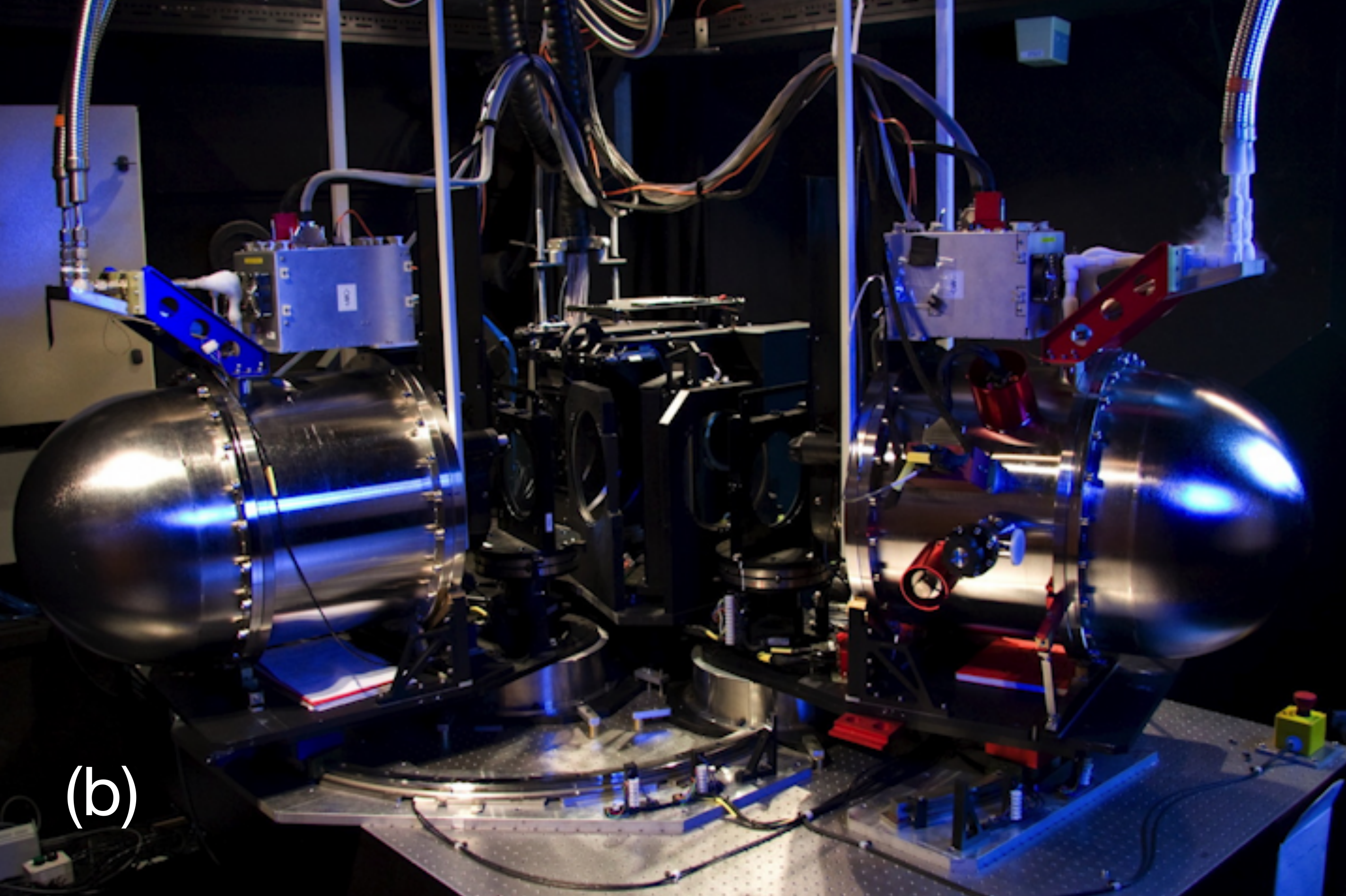}
  \end{center}
  \caption{\label{fig:AAOmega}(a)~A schematic of the AAOmega
    spectrograph showing the main components of the instrument. (b)~A
    photo of the AAOmega spectrograph in operation on the AAT.}
\end{figure}

After the conclusion of the 2dFGRS, the original 2dF spectrographs were
replaced with the AAOmega double-beam spectrograph\cite{Saunders2004,
  Smith2004, Sharp2006}. AAOmega is a bench-mounted system an f/3.15
Schmidt collimator, a dichroic beam-splitter, volume phase holographic
(VPH) gratings, and articulated f/1.3 Schmidt cameras (see
Figure~\ref{fig:AAOmega}). It accommodates 392 fibers and covers the
wavelength range 370\,nm to 950\,nm at spectral resolutions from
$R$=1000 to $R$=7500. It is floor-mounted in a thermally isolated
environment with a fiber cable running 38m to the AAT's prime focus.
Despite the long fibre run, AAOmega achieves a throughput of
approximately 20\% in both the blue and red arms, a gain of more than a
factor of 2 over the 2dF spectrographs. This improvement is due to a
number of factors: the use of highly efficient VPH gratings and optical
coatings, and to new higher-performance CCDs. AAOmega also improved on
2dF by a factor of 2 in resolution, while the spectral stability is an
order of magnitude better. AAOmega commenced science observations at the
AAT in early 2006 and is still in high demand today, being used with the
(refurbished) 2dF fiber positioning system, the KOALA\cite{Ellis2012,
  Zhelem2014} wide-field IFU feed, and the SAMI\cite{Croom2012,
  Bryant2015} multi-IFU system.

AAOmega has been used for a number of galaxy surveys, including the
ongoing GAMA\cite{Driver2011} and SAMI\cite{Bryant2015, Allen2015}
surveys of galaxy properties and evolution, and the
2SLAQ\cite{Cannon2006}, Wigglez\cite{Drinkwater2010}, and ongoing
OzDES\cite{Yuan2015} surveys which had cosmological goals. Of the
cosmological surveys using AAOmega, WiggleZ, which focussed on the
evolution of the geometry of the universe and the growth of structure at
redshifts up to $z \approx 1$, has had the greatest impact.

WiggleZ\cite{Drinkwater2010, Parkinson2012} measured redshifts for
238,000 galaxies starforming galaxies with $0.2 \lesssim z \lesssim 1$
($z_{median}$=0.6) in 7 regions covering approximately 1000\,deg$^2$ of
sky with a total volume of $\sim$1\,Gpc$^3$. The main results from the
WiggleZ survey were:
(a)~using the BAO standard ruler to measure the geometry of the
  universe, by mapping the distance-redshift relation and measuring the
  cosmic expansion history by combining the Alcock-Paczynski test and distant
  supernovae\cite{Blake2011b, Blake2011c, Blake2011d, Kazin2014};
(b)~measuring the growth rate of large-scale structure using redshift
  space distortions of the two- and three-point galaxy correlation
  functions\cite{Blake2010, Blake2011a, Contreras2013, Marin2013};
(c)~jointly measuring the expansion and growth history out to 
  $z \approx 1$\cite{Blake2012}; 
(d)~constraining the sum of the neutrino masses\cite{RiemerSorensen2012};
(e)~tracking the transition to large-scale cosmic
  homogeneity\cite{Scrimgeour2012}; and
(f)~fitting the galaxy power spectrum to determine six key
  cosmological parameters ($\Omega_{\rm b}h^2$, $\Omega_{\rm CDM}$,
  $H_0$, $\tau$, $A_{\rm s}$, $n_{\rm s}$) and five supplementary
  parameters ($n_{\rm run}$, $r$, $w$, $\Omega_{\rm k}$, $\sum m_\nu$),
  and showing that all are consistent with the standard $\Lambda$CDM
  model\cite{Parkinson2012}.

\subsection{BOSS \& eBOSS}
\label{sec:BOSS}

\begin{figure}
  \begin{center}
    \includegraphics[width=\textwidth]{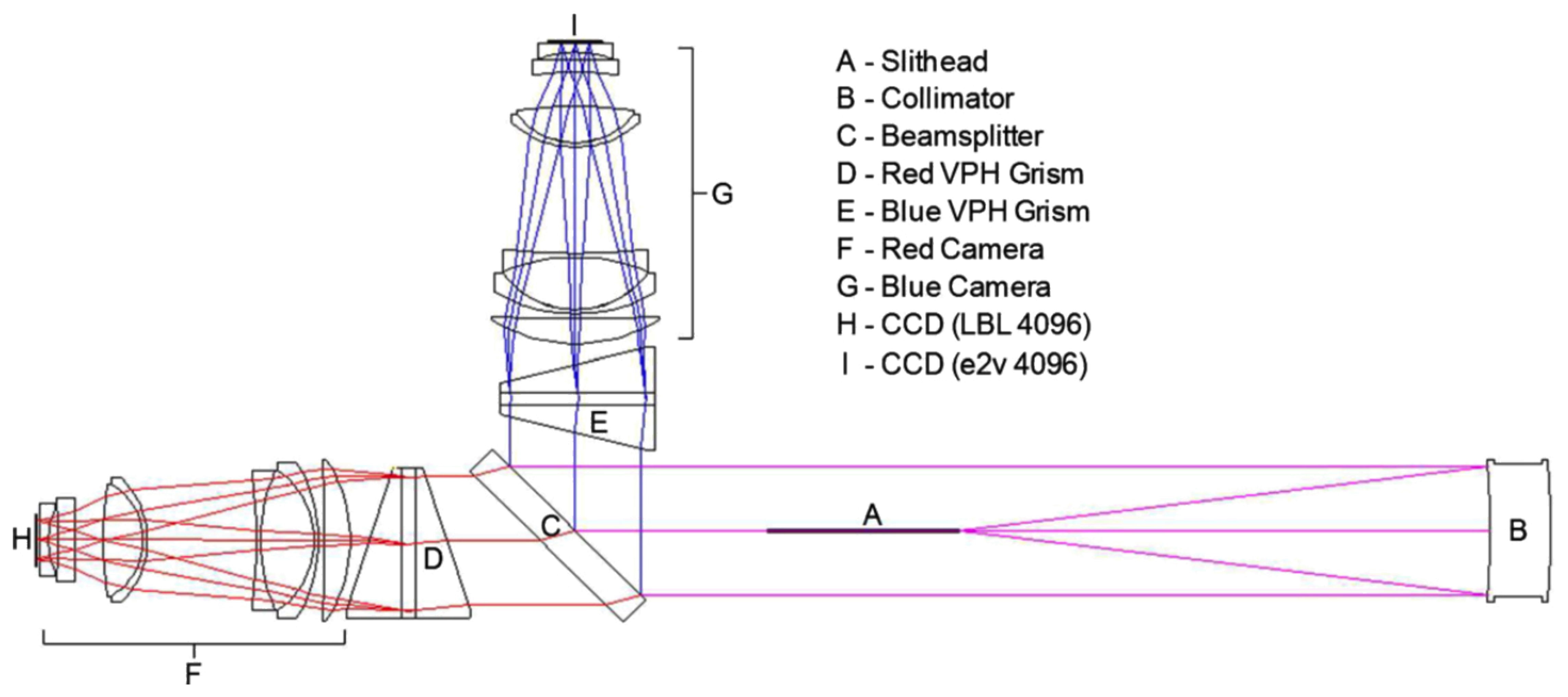}
  \end{center}
  \caption{\label{fig:BOSSinst}The optical layout for the upgraded SDSS
    spectrographs used for the BOSS and eBOSS surveys\cite{Smee2013}.}
\end{figure}

The current state-of-the-art in precision cosmology MOS surveys is
represented by the recently completed Baryon Oscillation Spectroscopic
Survey\cite{Eisenstein2011, Ahn2012, Dawson2013, Alam2015} (BOSS, part
of SDSS-III) and the ongoing extended BOSS\cite{Dawson2016} (eBOSS, part
of SDSS-IV). These surveys were enabled by a 2009 upgrade to the
original SDSS spectrographs\cite{Smee2013} on the Apache Point
Observatory 2.5-metre telescope, just as WiggleZ and other AAT surveys
were enabled by AAOmega replacing the original 2dF spectrographs. As
with AAOmega, the upgrade to the SDSS spectrographs involved volume
phase holographic gratings and more modern CCD detectors, and improved
the peak efficiency by nearly a factor of 2, while extending the
spectral range to 360--1000\,nm and increasing the multiplex from 640 to
1000 fibers (500 per spectrograph). To achieve the higher multiplex,
retain the spectral sampling given the smaller CCD pixels, and match the
source size for the more distant BOSS galaxies, the fiber diameter was
reduced from 3\,arcsec to 2\,arcsec. The overall layout of the upgraded
spectrographs is the same as the original SDSS spectrographs (see
Figure~\ref{fig:SDSSinst}) but with a revised central optics assembly,
as illustrated in Figure~\ref{fig:BOSSinst}. As with the SAMI multi-IFU
feed for AAOmega, the new SDSS spectrographs have been provided with
multi-IFU capability as part of the MaNGA survey\cite{Bundy2015} that is
obtaining spatially resolved spectroscopy for $\sim$10,000 galaxies.

BOSS observations were performed from 2008 to 2014, starting with
imaging in 2008 and spectroscopy from 2009, when the upgraded SDSS
spectrographs became available. The final data release\cite{Alam2015}
(SDSS DR12) was made public in 2015, and contained spectra for
1.37~million unique galaxies and QSOs, made up of 862,735 galaxies from
the LOWZ target catalogue ($0.15<z<0.4$), 343,160 galaxies from the
CMASS target catalogue ($0.4<z<0.8$), and 158,917 QSOs with $2.15<z<3.5$
used to study large-scale structure in the Lyman-$alpha$ forest. The
BOSS spectroscopy covers an effective area of 9376\,deg$^2$ over two
large contiguous regions in the north and south Galactic caps.

\begin{figure}
  \begin{center}
    \includegraphics[width=0.48\textwidth]{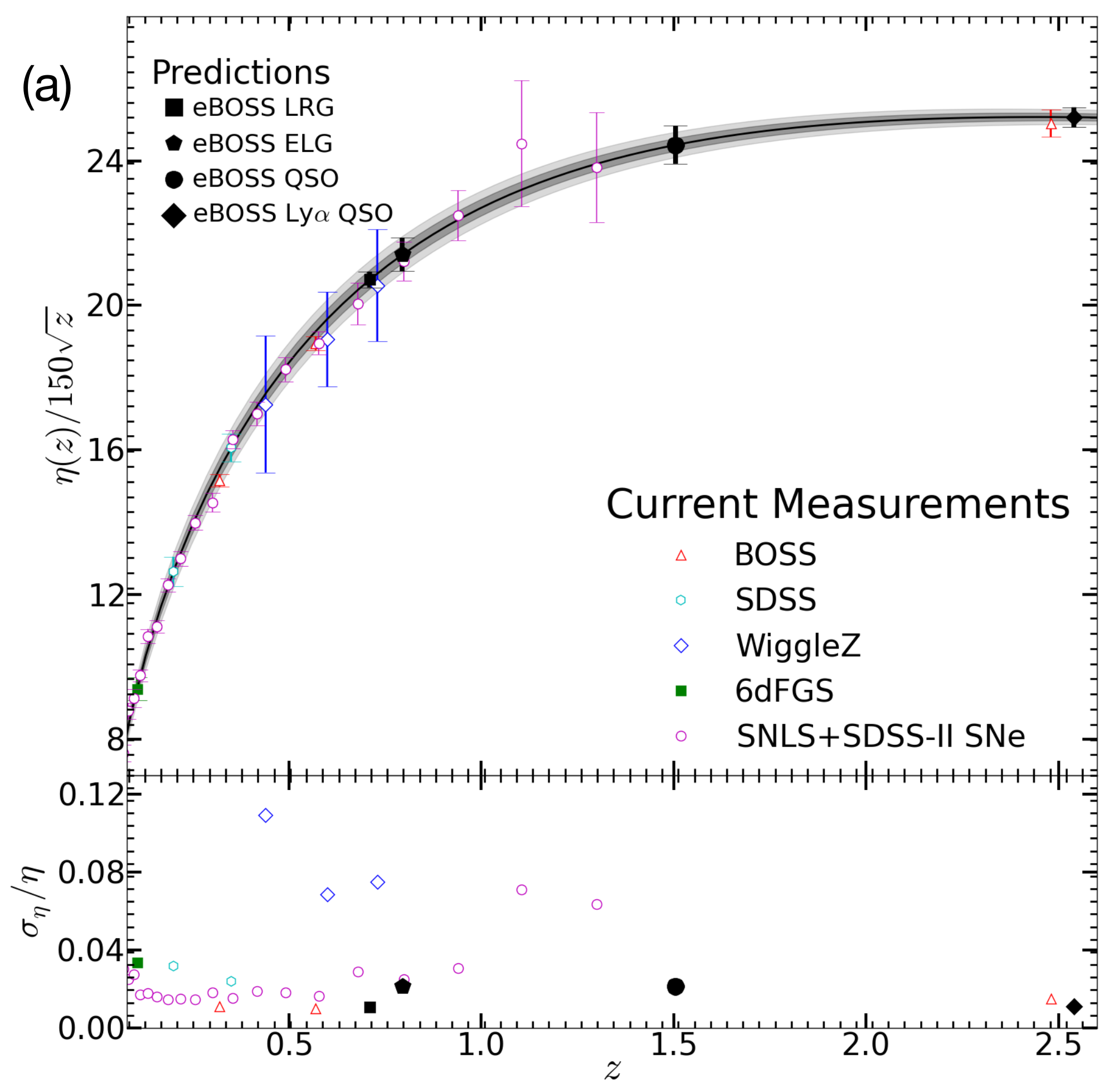}
    \includegraphics[width=0.48\textwidth]{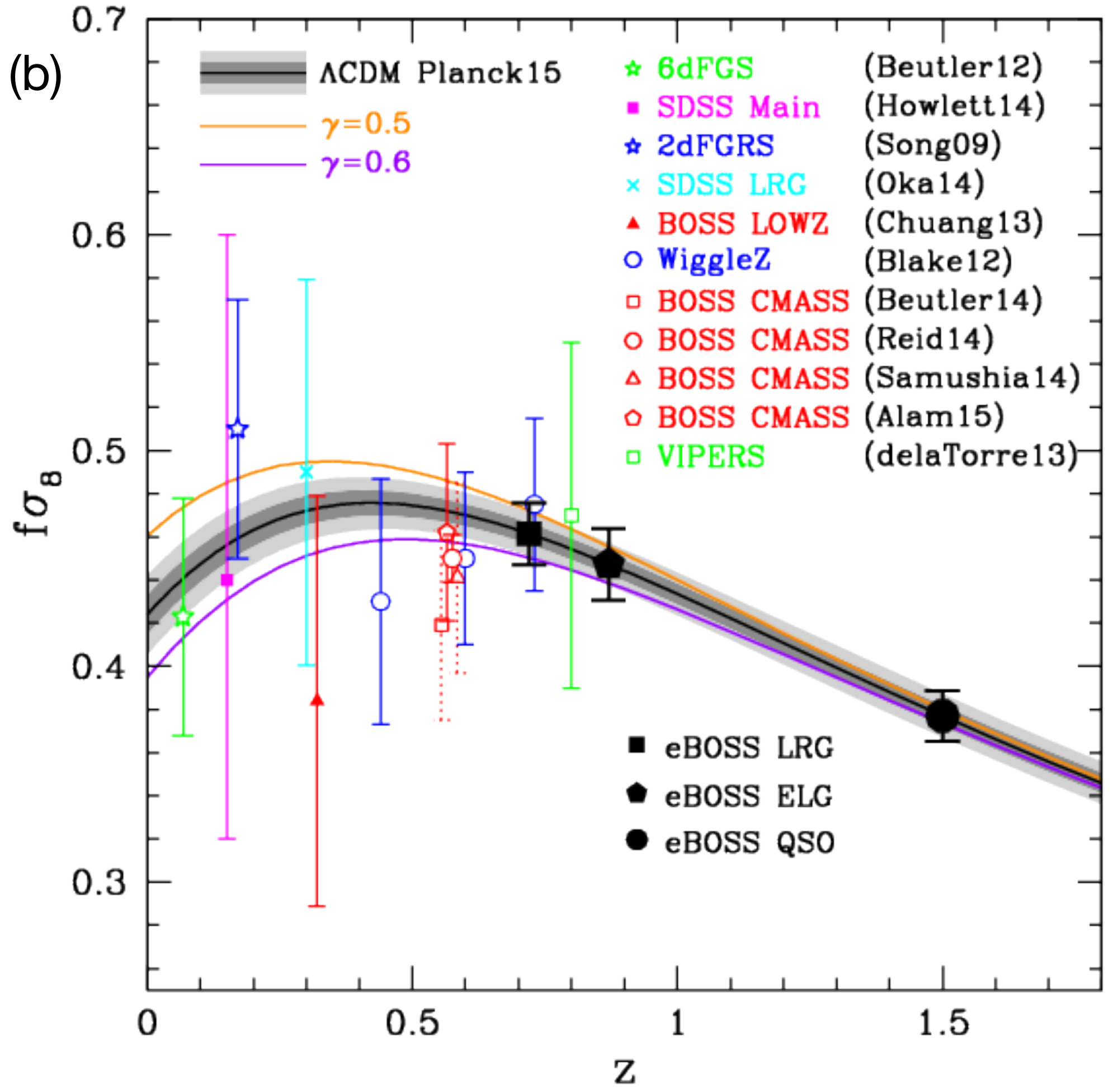}
  \end{center}
  \caption{\label{fig:BOSSresults}Measurements of the distance-redshift
    relation and the growth rate of structure from MOS
    surveys\cite{Dawson2016}. (a)~The comoving distance $\eta(z)$ versus
    redshfit $z$ relation predicted from the Planck $\Lambda$CDM model
    (shown as the solid black curve, with 1$\sigma$ and 2$\sigma$
    uncertainties in gray); the existing measurements from BAO surveys
    (6dFGS, SDSS, WiggleZ and BOSS); the expected measurements from
    eBOSS; and a compilation of SNe Ia measurements. (b)~The growth rate
    of structure $f\sigma_8$ versus redshfit $z$ relation predicted from
    the Planck $\Lambda$CDM model (shown as the solid black curve, with
    1$\sigma$ and 2$\sigma$ uncertainties in gray); the existing
    measurements from previous redshift-space distortions from redshift
    surveys (6dFGS, 2dFGRS, SDSS, WiggleZ, VIPERS and BOSS); and the
    expected measurements from eBOSS. Also shown are curves
    corresponding to theories of gravity predicting growth going as
    $f=\Omega^\gamma$, with $\gamma$ differing slightly from the General
    Relativity value of 0.55.}
\end{figure}

Some of the main cosmological results of the BOSS survey, and relevant
preceding surveys, are summarized\cite{Dawson2016} in
Figure~\ref{fig:BOSSresults}. The lefthand panel shows predictions and
observations of the relation between comoving distance and redshift,
which constrains the equation of state of the universe. The prediction
comes from the Planck\cite{Planck2015a} $\Lambda$CDM model, while the
observed points are from previous BAO survey measurements by
6dFGS\cite{Beutler2011}, WiggleZ\cite{Parkinson2012} and
SDSS/BOSS\cite{Xu2013, Ross2015}, and a compilation of current SNe Ia
measurements\cite{Betoule2014}. The righthand panel of
Figure~\ref{fig:BOSSresults} shows the predicted growth rate of
structure ($f\sigma_8$) as a function of redshift, and compares the
Planck\cite{Planck2015a} $\Lambda$CDM model to the measurements based on
redshift-space distortions (RSD) from 6dFGS\cite{Beutler2012},
2dFGRS\cite{Cole2005, Song2009}, SDSS LRGs\cite{Oka2014},
WiggleZ\cite{Blake2012}, VIPERS\cite{delaTorre2013} and
BOSS\cite{Chuang2013, Beutler2014a, Reid2014, Samushia2014, Alam2015}.
Because the growth rate simultaneously tests both the cosmological model
and the theory of gravity, the panel alsos shows the predictions for
theories of gravity predicting growth going as $f=\Omega^\gamma$, with
$\gamma$ differing slightly from the General Relativity value of 0.55,
demonstrating the sensitivity of these measurements to such
alternatives.

All existing measurements are consistent with the standard $\Lambda$CDM
model at the 5--10\% level; the next generation of MOS surveys aim to
tighten these constraints to a level approaching 1\%. The first of the
new surveys is eBOSS\cite{Dawson2016, Zhao2016}, a 6-year survey (part
of the SDSS-IV program) that started in 2014 and plans to complement and
extend BOSS. It has the goal of measuring the distance-redshift relation
from the BAO ruler with a precision of a few percent in each of four
redshift bins over the range $0.6<z<2.2$. It will use four different
tracers to cover this range: 250,000 luminous red galaxies with median
redshift $z \approx 0.7$; 195,000 emission line galaxies with median
redshift $z \approx 0.9$; 500,000 QSOs over $0.9<z<2.2$; and
Lyman-$\alpha$ forest measurements using 120,000 QSOs at $z>2.1$. As
well as determining the evolution of the geometry of the universe in
order to determine the equation of state of dark energy, eBOSS aims to
make stronger tests of General Relativity on cosmological scales through
redshift-space distortion measurements, look for evidence of
non-Gaussianity in the primordial density field, and tighten the
constraints on the sum of the masses of all neutrino species. The
constraints that eBOSS is predicted to yield\cite{Dawson2016}on the
comoving distance and the growth rate of structure as functions of
redshift (assuming the Planck\cite{Planck2015a} $\Lambda$CDM model) are
shown in Figure~\ref{fig:BOSSresults}.

eBOSS is currently the state of the art as far as MOS cosmological
surveys are concerned. However there are, of course, plans afoot for
even more potent MOS surveys providing still greater discrimination
between cosmological models and yet higher precision in determining
cosmological parameters.

\section{THE FUTURE}
\label{sec:future}

\begin{figure}
  \begin{center}
    \includegraphics[width=0.48\textwidth]{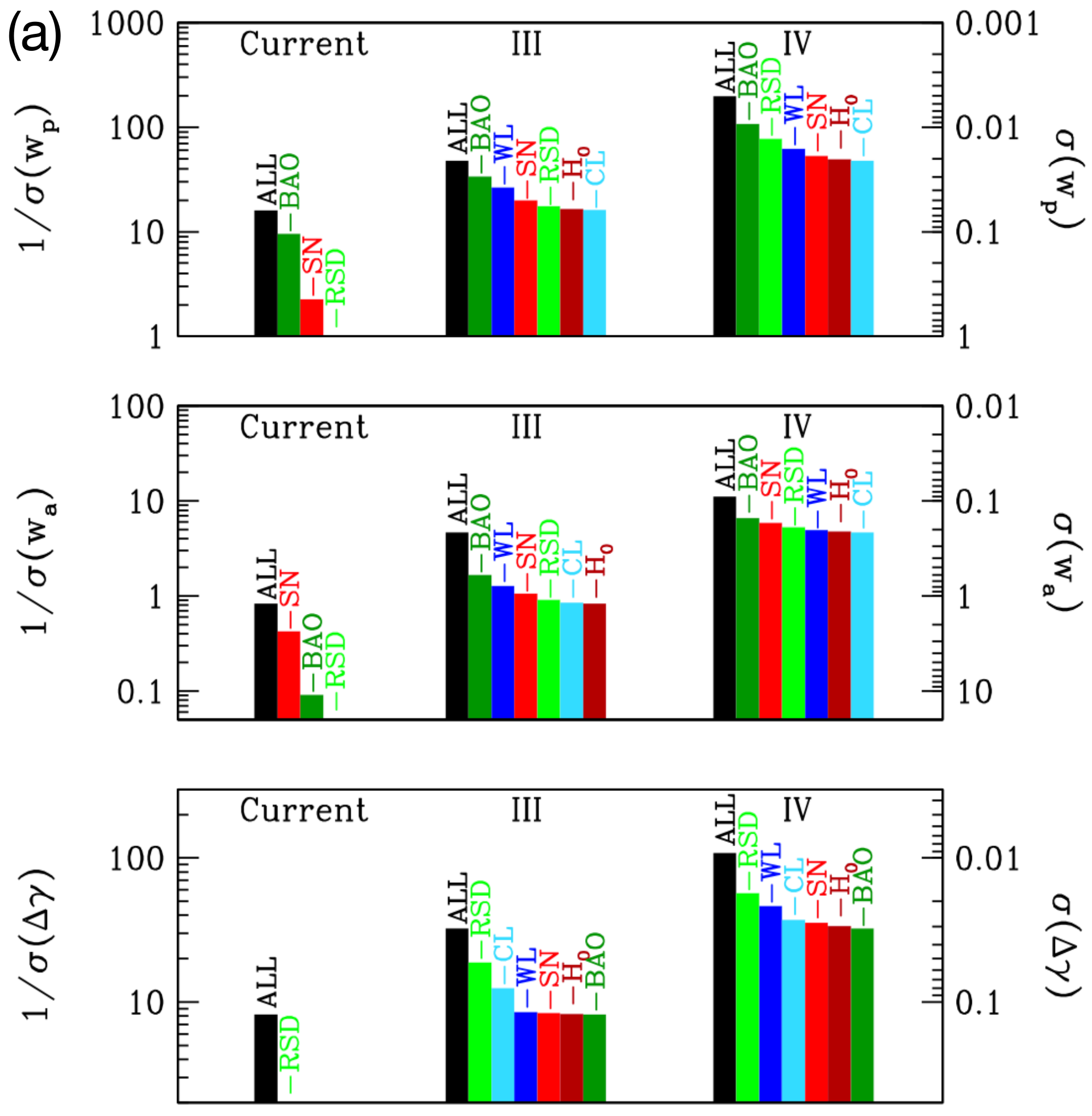}
    \includegraphics[width=0.48\textwidth]{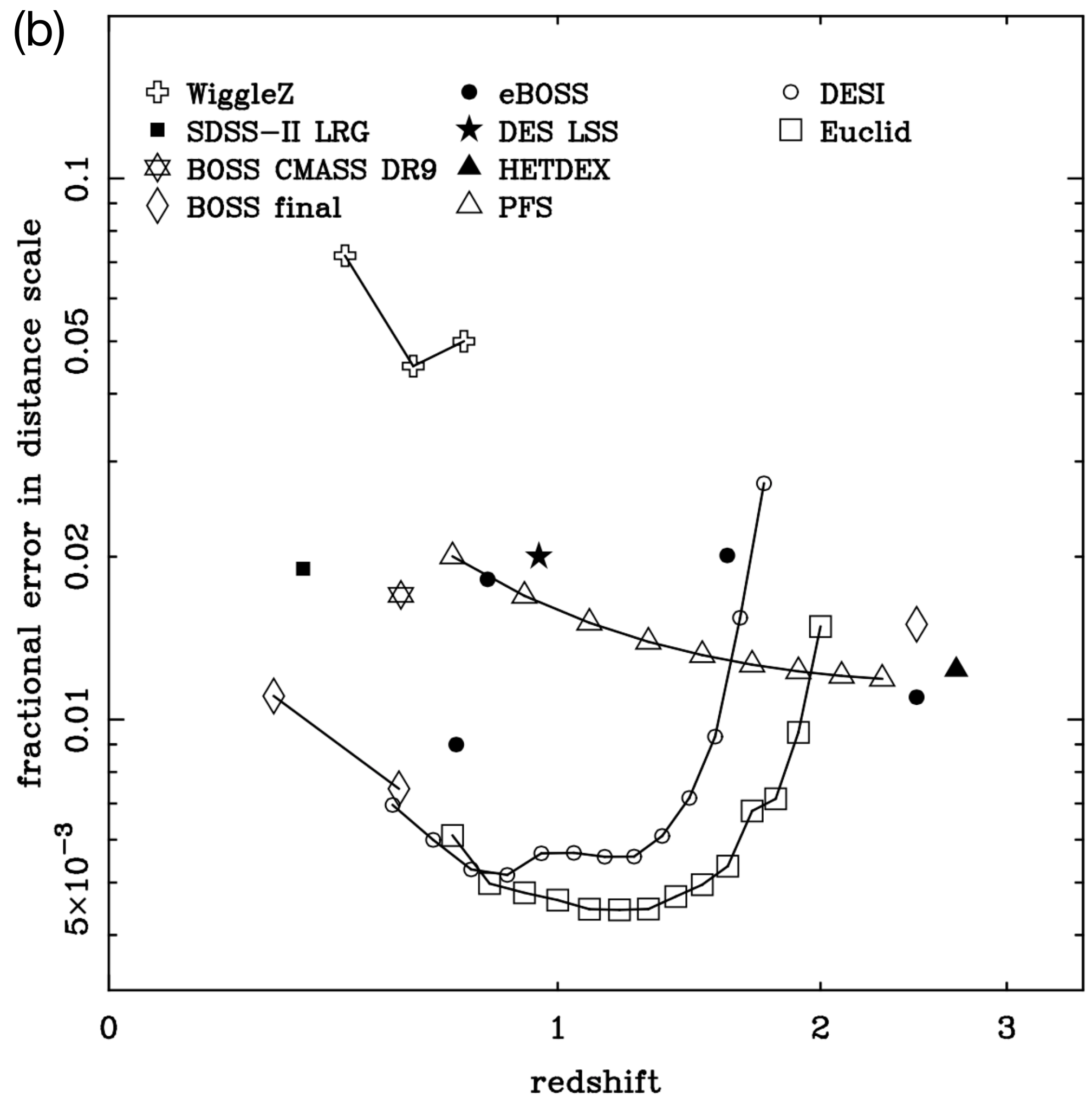}
  \end{center}
  \caption{\label{fig:DEconstraints}Current and predicted constraints
    from cosmological MOS surveys. (a)~Potential improvements in figures
    of merit for constraints on the dark energy equation of state and
    the theory of gravity for existing and potential future BAO and RSD
    redshift surveys, and other methods\cite{Weinberg2013a}. (b)~The
    fractional error in constraints on the co-moving distance versus
    redshift relation, at various redshifts, from current and future MOS
    surveys\cite{Percival2013}.}
\end{figure}

There are a number of proposed new cosmological MOS surveys planned for
the near future, plus a wider variety of surveys aiming to explore the
high-redshift universe using the next-generation Extremely Large
Telescopes. The general program for these surveys is to continue the
push for ever-larger samples over ever-larger volumes covering a wider
range of redshfits. The focus of such surveys is ever-more-precise
constraints on cosmological parameters, with a particular emphases on
testing the nature of dark energy and the theory of gravity. Although
all observations to date of the cosmic expansion history and the growth
of structure are consistent with a flat $\Lambda$CDM$+$GR model with
$\Omega_M \approx 0.3$ and $\Omega_\Lambda \approx 0.7$, a number of
plausible alternative models are also consistent with existing data.
However BAO and RSD have to the potential, for sufficiently large
surveys, to provide $\sim$10$\times$ stronger constraints on the
equation of state of dark energy and the nature of the gravitational
force. This opportunity is quantified by improvement in the relevant
figures of merit, expressed as the inverse of the uncertainties on the
$w_a$ and $w_p$ parameterisations of the DE equation of state and
deviations $\Delta\gamma$ from the GR value of $\gamma = 0.55$. These
improvements for BAO and RSD measurements are shown for possible future
MOS surveys\cite{Weinberg2013a} in Figure~\ref{fig:DEconstraints}a; another
view\cite{Percival2013} of the constraints, based on the actual and
predicted fractional errors in the distance--redshift relation for
existing and future MOS surveys, is provided in
Figure~\ref{fig:DEconstraints}b.

\subsection{Low Redshift Surveys}
\label{sec:lowz}

At low redshifts, the Taipan survey\cite{Colless2013} on the
newly-refurbished 1.2-metre UK Schmidt Telescope (UKST) aims to obtain
high-precision constraints on the $z\approx0$ values of key parameters
such as the Hubble constant ($H_0$) and the growth rate of structure.
The survey will use the new TAIPAN\cite{Kuehn2014} fibre positioner and
spectrograph, which are currently under construction at the Australian
Astronomical Observatory. The TAIPAN positioner is a prototype for the
MANIFEST\cite{Saunders2010, Goodwin2012, Lawrence2014} fibre positioning
system planned for the Giant Magellan Telescope (GMT), and uses
autonomous piezo-electric micro-robots (`starbugs') to move the optical
fibres about in the curved focal plane of the UKST. The TAIPAN
spectrograph is a fixed-format, two-channel spectrograph covering the
range 370-850\,nm at R=2000-2400 and delivering velocity resolution
$\sigma_v \approx 60$\,km\,s$^{-1}$ with a total efficiency of $\sim$30\%.

\begin{figure}
  \begin{center}
    \includegraphics[width=\textwidth]{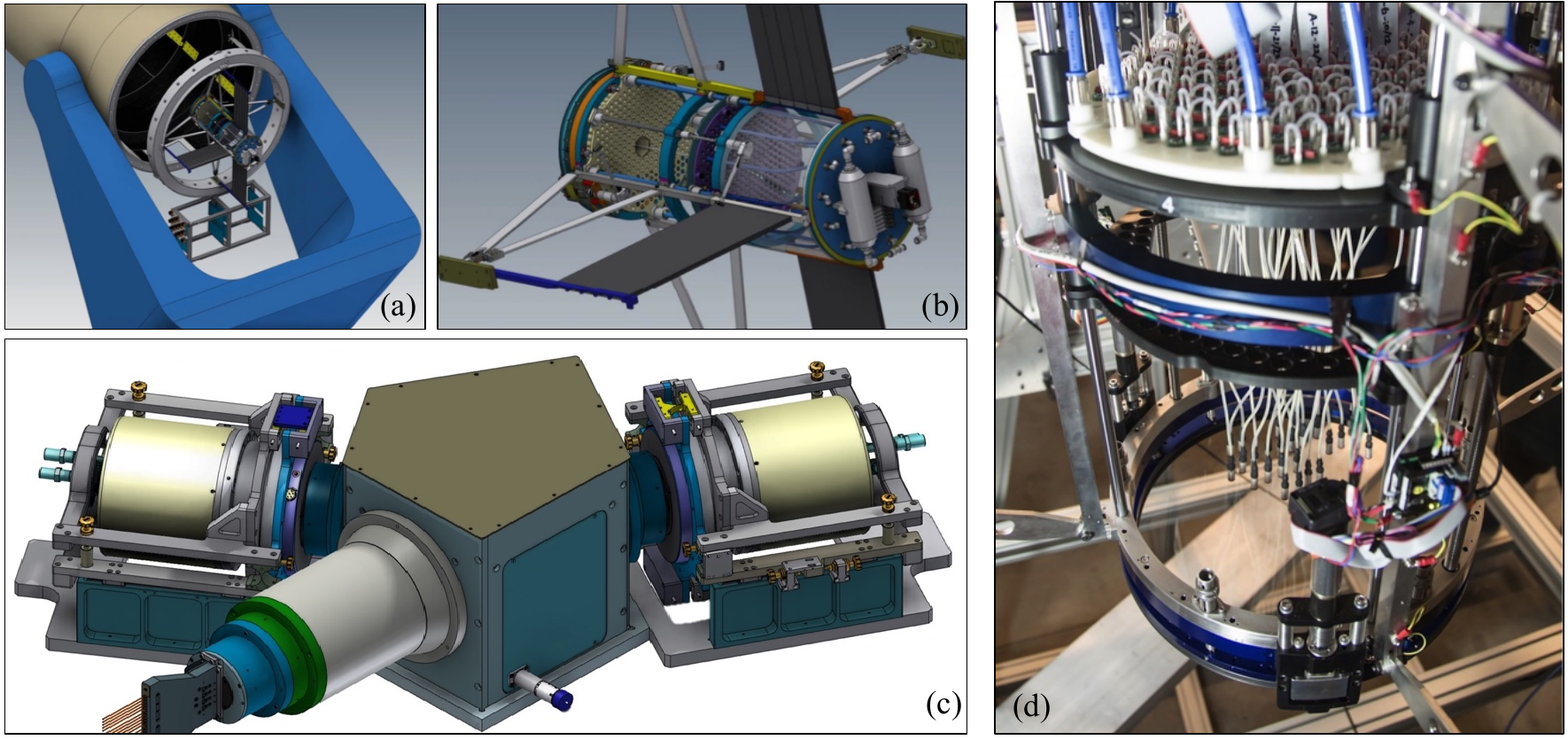}
  \end{center}
  \caption{\label{fig:TaipanSystem}The components of the UKST+TAIPAN
    system: (a)~the location of the TAIPAN fibre positioner within the
    UKST (in cutaway view); (b)~a close-up rendering of the fibre
    positioner system containing the Starbugs; (c)~the TAIPAN
    spectrograph layout, showing the blue and red channels covering the
    whole optical wavelength range; and (d)~an image of some Starbugs
    being tested during the manufacturing process at AAO.}
\end{figure}

Covering essentially no redshift range, the Taipan survey cannot examine
the evolution of such quantities, but it can measure them in the
present-day universe without extrapolations from high redshift that
require assumptions about the cosmological model. For example, the
Planck\cite{Planck2015a} CMB observations yield a measurement of
$H_0 = 67.3 \pm 0.7$\,km\,s$^{-1}$\,Mpc$^{-1}$, but this assumes a
$\Lambda$CDM model with particular parameters in order to transform
measurements made at $z \approx 1100$ to parameters at $z = 0$. The
Taipan survey, by contrast, will obtain redshifts for $\sim$1~million
galaxies at redshifts $z < 0.2$ and directly measure $H_{0.1}$ with 1\%
precision from the BAO standard ruler that is imprinted in the
large-scale structure. Taipan thus will provide a bookend measurement of
the present-day expansion rate with a precision matching that of
Planck's measurement of the expansion rate shortly after the Big Bang
and that of ongoing and future cosmological surveys, such as the eBOSS
and the Dark Energy Survey (DES), at high and intermediate redshifts. A
recent review of the prospects in this field\cite{Suyu2012} stated that:
“A measurement of the local value of $H_0$ to 1\% precision and accuracy
would provide key new insights into fundamental physics questions and
lead to potentially revolutionary discoveries.” While ongoing
programs\cite{Riess2016} to measure $H_0$ at low redshift using
supernovae as standard candles aim for a similar level of precision to
Taipan, they have different and (arguably) greater systematic errors
with which they must contend.

Similarly, Taipan is expected to obtain substantially better
measurements at $z \approx 0$ of the growth rate of structure
($f\sigma_8$) and the velocity field scaling parameter ($\beta$), both
of which should be determined to better than 5\% precision. These
low-redshift measurements are important because, again, they provide
bookends to CMB measurements, but also because they depend on both the
cosmological model and the theory of gravity, allowing tests of general
relativity against alternative models. Combining Taipan's low-redshift
constraints with higher-redshift MOS surveys and the CMB observations
will significantly tighten the constraints on all important extensions
to the standard cosmological model---in particular: the nature of dark
energy and its evolution with time, the curvature of the universe as a
test of inflationary models, the mass of neutrinos, and the total number
of families of relativistic particles\cite{Riess2011, Weinberg2013a}.

\subsection{Higher Redshift Surveys}
\label{sec:medz}

The primary focus of cosmological MOS surveys at higher redshifts is
tracing the evolution of the geometry of the universe and the growth of
large-scale structure over a period encompassing the epoch during which
the universe changed from being matter-dominated to
dark-energy-dominated (this occurs at $z \approx 0.8$). For direct
redshift measurements, the upper limit on this range is set by the depth
achievable with 4-metre and 8-metre telescopes under the constraint that
for each redshift bin the survey must cover at least $\sim$1\,Gpc$^3$ in
order to determine the BAO signal, and hence the co-moving distance at
that redshift, with sufficient precision to usefully constrain the dark
energy equation of state. The most ambitious such survey currently
planned is DESI\cite{Levi2013, DESI-FDR-I_2016}, which aims to provide
at least an order of magnitude improvement over BOSS/eBOSS, both in the
comoving volume of the universe probed and in the total number of
galaxies mapped. DESI is planned to start in 2018, with observations
running for 5 years. It will be complementary to other planned new
cosmological surveys such as those using the Large Synoptic Survey
Telescope (LSST) and the EUCLID and WFIRST satellite missions.

\begin{figure}
  \begin{center}
    \includegraphics[width=0.8\textwidth]{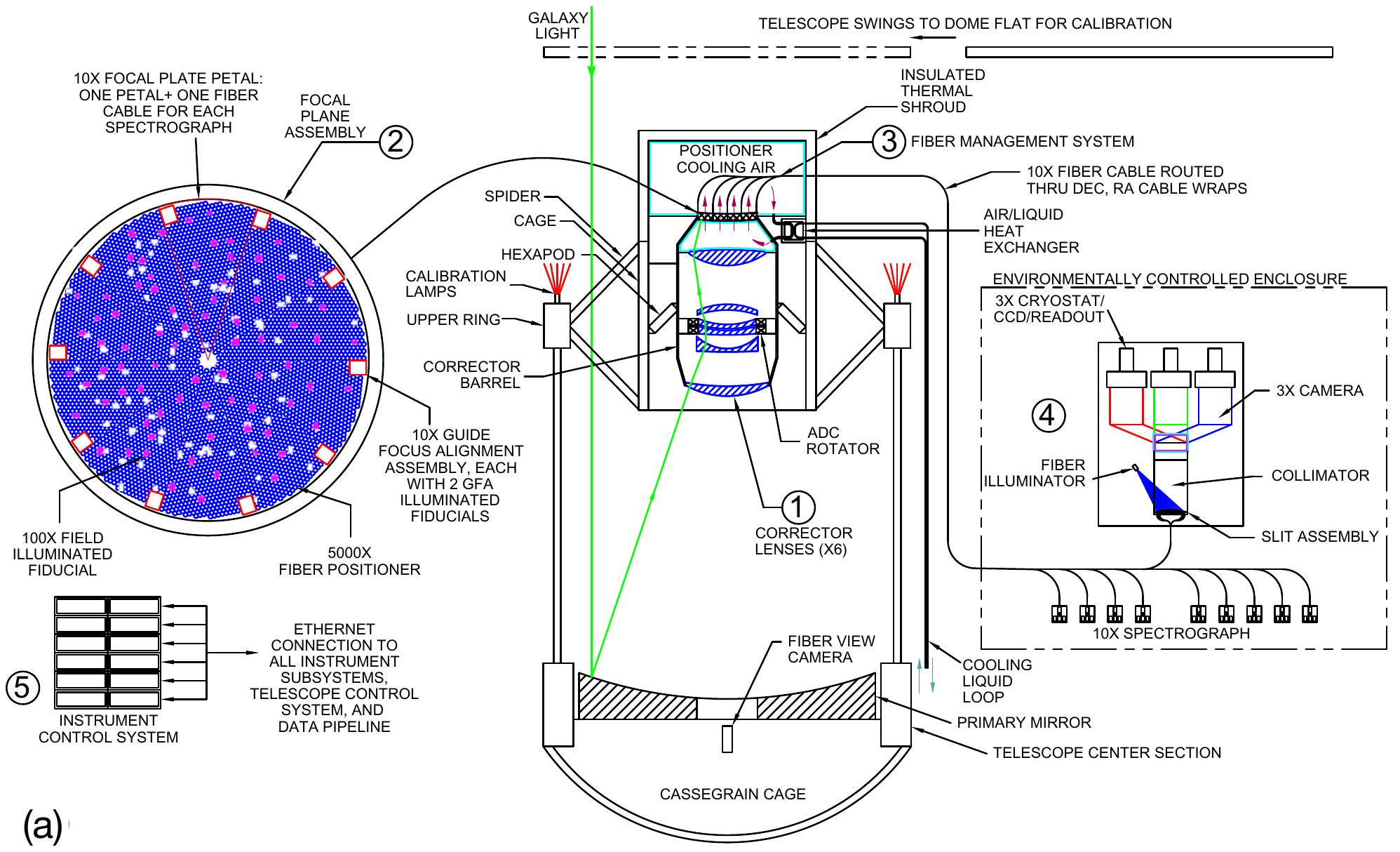} ~ 
    \includegraphics[width=0.065\textwidth]{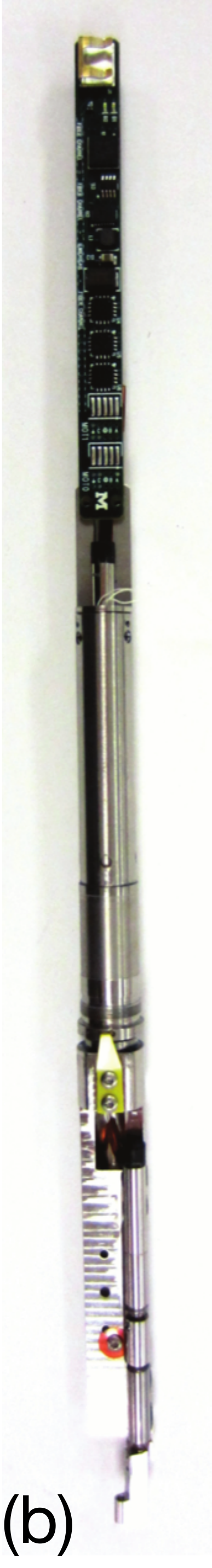}
  \end{center}
  \caption{\label{fig:DESIinstrument}The DESI
    instrument.\cite{DESI-FDR-II_2016} (a)~Block diagram showing the
    main components of the instrument on the KPNO Mayall 4-metre
    telescope: 1.~wide-field corrector and atmospheric dispersion
    compensator producing a 3.2-degree field of view at the telescope
    prime focus; 2.~focal plane assembly, with 5000 fibers in 10 'petal'
    segments covering the field of view; 3.~fiber management system and
    fiber cable run; 4.~ten spectrographs, each taking 500 fibers from
    one of the 'petals' and each with three (blue/green/red) arms.
    (b)~one of the robotic actuators for positioning individual fibres.}
\end{figure}

The DESI instrument\cite{DESI-FDR-II_2016} will be a fiber-fed
spectrograph using a robotic positioner system and will be capable of
taking up to 5000 simultaneous spectra over the range 360--980\,nm. The
main components of the system, and one of the fiber positioners, are
shown in Figure~\ref{fig:DESIinstrument}. The instrument will be
installed at the prime focus of the 4-metre Mayall telescope at Kitt
Peak and requires a new wide-field corrector and an atmospheric
disperion compensator, which produce a 3.2-degree diameter field of view
with an average scale of 14.1\,arcsec/mm. The 5000 fibers can be
positioned over 7.5\,deg$^2$ of the available 8.0\,deg$^2$ in the field
of view. The fiber density in the focal plane is 667\,deg$^{-2}$ and the
individual fibers have 107\,$\mu$m (1.5\,arcsec) cores. The positioners
are arrayed hexagonally, with a 10.4\,mm (147\,arcsec) pitch between
fibers. Each positioner has two rotational degrees of freedom allowing
it to reach any point within a 6\,mm (85\,arcsec) radius. The focal
plane is divided into ten pie-slice-shaped petals containing 500 fibres,
each feeding one of the ten spectrographs. Volume phase holographic
gratings provide spectral dispersion in the three spectrograph channels
(360--593\,nm, 566--772\,nm and 747--980\,nm) with resolutions greater
than 2000, 3200, and 4100 respectively. The blue arm of the
spectrographs use CCDs from Imaging Technologies Lab, while the red and
near-infrared channels use LBNL CCDs; all will be 4k$\times$4k with
15\,$\mu$m pixels. The system will have overall throughput around
35--40\% over most of the spectral range, except at the bluest
wavelengths.

\begin{figure}
  \begin{center}
    \includegraphics[width=0.63\textwidth]{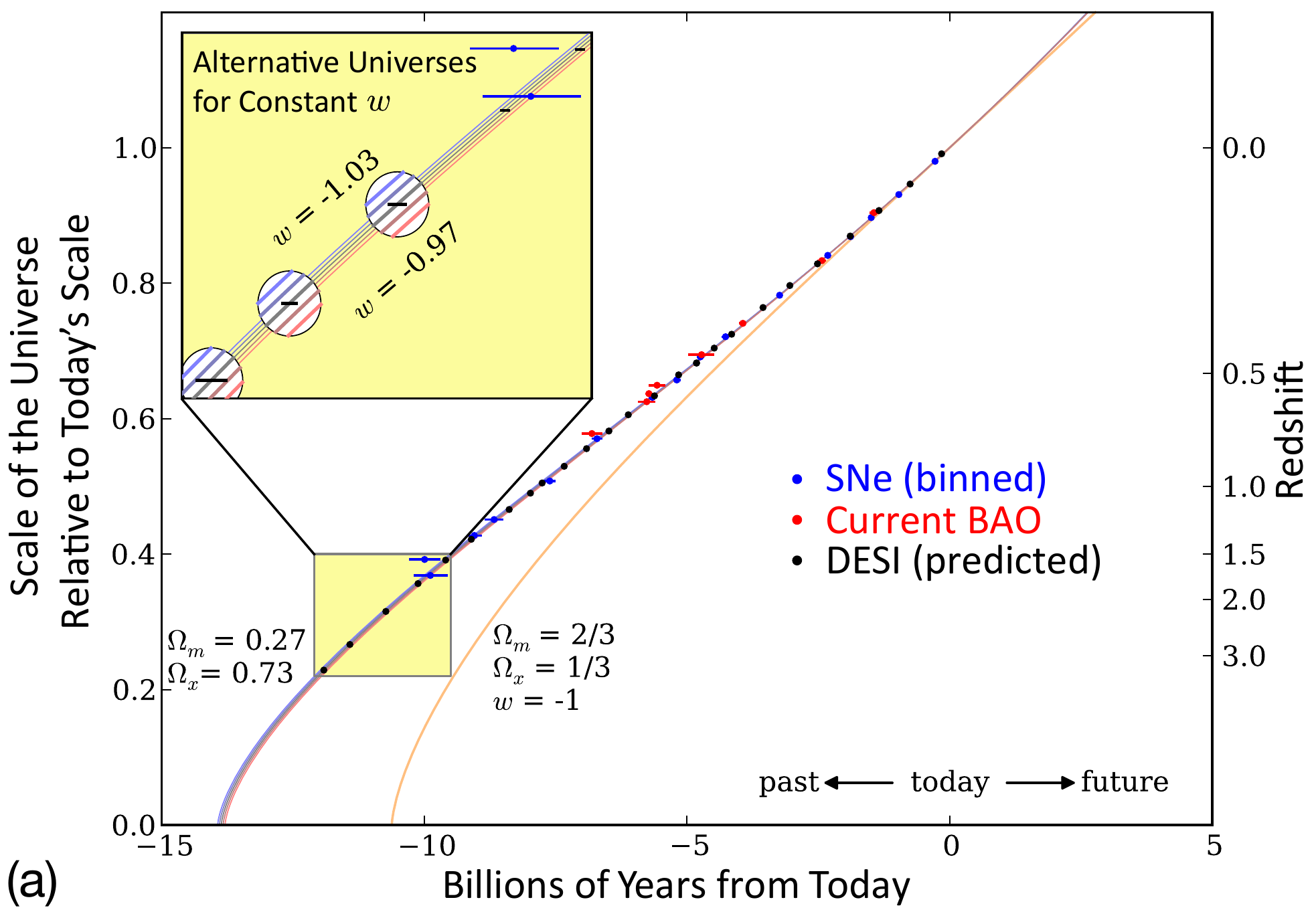}
    \includegraphics[width=0.34\textwidth]{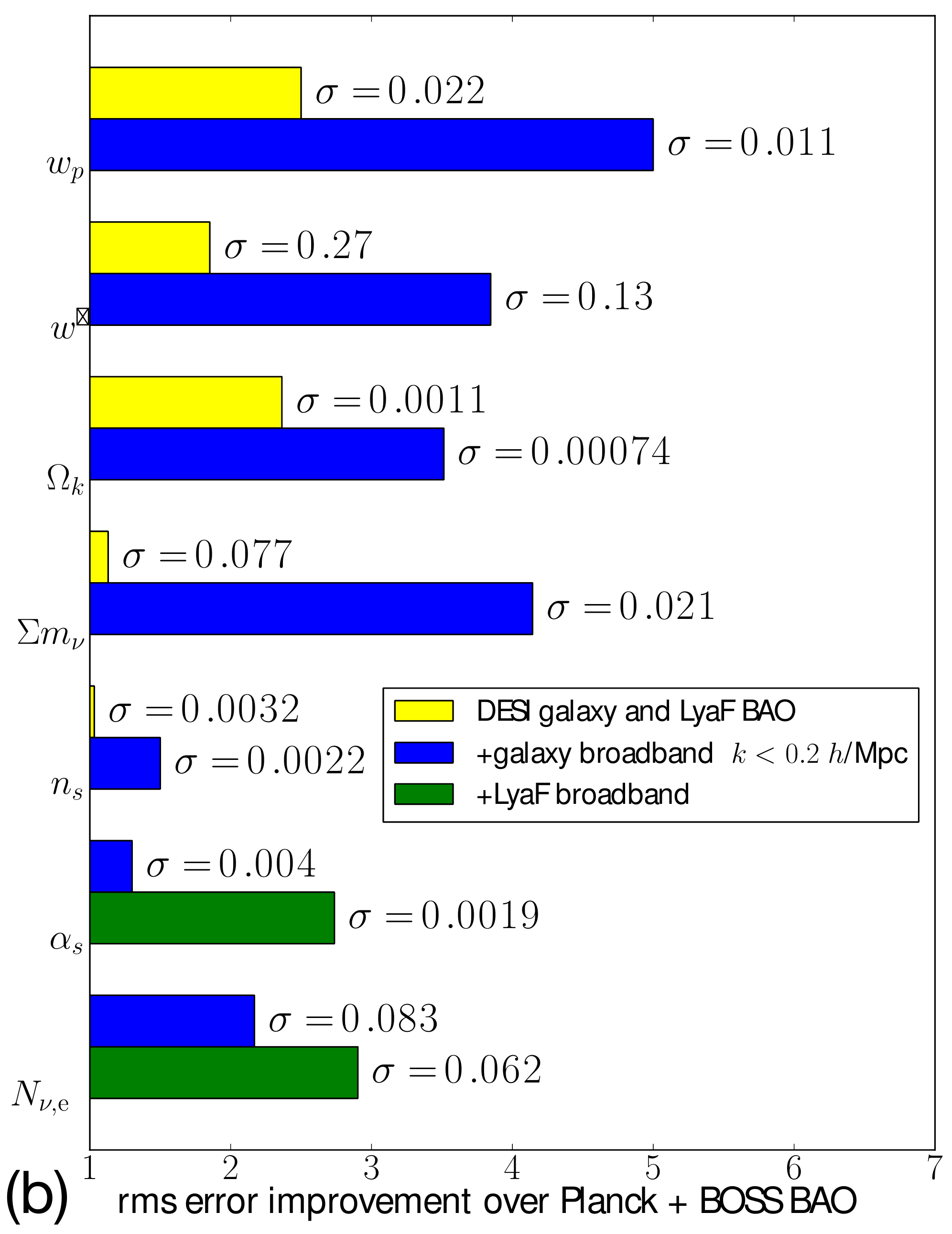}
  \end{center}
  \caption{\label{fig:DESIconstraints}Predicted cosmological constraints
    and figures of merit for the DESI survey.\cite{DESI-FDR-I_2016}
    (a)~Measurements of the expansion rate, comparing DESI to the best
    current BAO and supernovae measurements. The inset illustrates the
    very high precision needed to distinguish models with constant $w$
    ranging from 0.97 to 1.03. (b)~The improvements predicted for DESI
    in determining $w_p$, $w^\prime = w_a$, $\Omega_k$, $\sum m_\nu$,
    $n_s$, $\alpha_s$, and $N_{\nu,e}$, the number of neutrino-like
    (relativistic) species..}
\end{figure}

The DESI instrument will be used to conduct a 5-year survey covering
14,000~deg$^2$ that will observe four classes of objects in order to map
the matter distribution over the widest possible range of
redshifts\cite{DESI-FDR-I_2016}. At moderate redshifts (up to
$z \approx 1$), DESI will observe luminous red galaxies (LRGs) in dark
time and a magnitude-limited sample of up to 10 million galaxies with
median redshift $z \approx 0.2$ in bright time. To map the matter
distribution at higher redshifts, DESI will use emission line galaxies
(ELGs, specifically those with strong [O\,II] lines) to reach
$z \approx 1.7$. At the highest redshifts, DESI will use QSOs both as
direct tracers of the matter distribution and, at $2.1 < z < 3.5$, as
probes of the Ly-α forest, tracing the distribution of neutral hydrogen.
In total, DESI aims to obtain more than 30~million galaxy and QSO
redshifts in order to measure the BAO feature and the matter power
spectrum, including redshift space distortions, over most of cosmic
history.

If DES achieves these observational goals, it will produce more than 30
separate measurements of the expansion rate of the universe, each with
precision better than 1\%, over the redshift range $z$=0.2 to $z$=3.5.
In terms of the Dark Energy Task Force figure of merit (FoM), which
measures the combined precision on the dark energy equation of state
today, $w_0$, and its evolution with redshift, $w_a$, DESI's galaxy BAO
measurements are predicted to achieve a FoM of 133, which is more than
3$\times$ better than the FoM for all previous galaxy BAO measurements
combined. If the Lyman-$\alpha$ forest measurements are also included,
then DESI's predicted FoM increases to 169. Finally, including galaxy
broadband power spectrum measurements for wavenumbers
$k < 0.1$\,h\,Mpc$^{−1}$, DESI's FoM rises still further to 332, and to
704 if it is possible to obtain reliable measurements for
$k < 0.2$\,h\,Mpc$^{−1}$.

In addition to providing vastly improved constraints on dark energy,
DESI will also measure the sum of neutrino masses with an uncertainty of
just 0.02\,eV (again, if the power spectrum is measured reliably for
$k < 0.2$\,h\,Mpc$^{−1}$). This should be sufficient to make the first
direct detection of the sum of the neutrino masses at $3\sigma$
significance and to rule out the the inverted mass hierarchy with 99\%
confidence (if the hierarchy is normal and the masses are minimal). The
survey will also place tighter constraints on alternative (non-GR)
theories of gravity and on models for inflation by measuring the
spectral index $n_s$ and its running with wavenumber, $\alpha_s$, and
also the velocity fields of in the infall regions around clusters of
galaxies. The cosmological constraints that DESI is predicted to
achieve, and the gains relative to existing results, are summarized in
Figure~\ref{fig:DESIconstraints}.

Other facilities on 4-metre and 8-metre telescopes that have the
capability to carry out similar surveys include: WEAVE\cite{Dalton2014}
on the William Herschel 4.2-metre telescope at La Palma Observatory
(1000 fibers over a 3\,deg$^2$ field from 2017); 4MOST\cite{deJong2014}
on ESO's 4.1-metre VISTA facility at Paranal (2400 fibers over a
4\,deg$^2$ field from 2021); and the Prime Focus
Spectrograph\cite{Takada2014} (PFS) on the Subaru 8.2-metre telescope on
Mauna Kea (2400 fibers over a 1.3\,deg$^2$ field from 2019). However
these facilities generally have much broader scientific goals than just
cosmology surveys, and often are general-use facilities rather than
dedicated survey facilities.

\subsection{ELT Surveys}
\label{sec:eltsurveys}

At higher redshifts, cosmological surveys such as BOSS and DESI use the
Lyman-$\alpha$ forest in quasar spectra to map large-scale structure and
place constraints on the geometry of the universe and the growth of
large-scale structure just as galaxy redshift surveys do at lower
redshifts. This approach will be even more powerful for the new
generation of Extremely Large Telescopes (ELTs). With collecting areas
6--15 times greater than the largest existing telescopes, ELTs can use
substantially fainter (and therefore much more common) background
sources such as Lyman-break galaxies to illuminate the Lyman-$\alpha$
forest. This means ELTs will be able to probe the matter distribution on
much shorter length scales than surveys using quasars and build much
larger samples. The disadvantage of ELTs is that they have significantly
smaller fields of view than the wide-field 4-metre telescopes (2--3\,deg
diameter) and 8-metre telescopes (up to 1.5\,deg diameter).

The largest ELT field of view is the 20~arcmin diameter provided by the
25-metre Giant Magellan Telescope (GMT). This field will be fully
exploited by the MANIFEST fiber system\cite{Saunders2010, Goodwin2012,
  Lawrence2014} (see also the paper by Lawrence et~al.\ in these
proceedings), which will use starbugs (as on the prototype TAIPAN system
on the UKST) to position hundreds of small integral field units
(effectively, image-slicers). MANIFEST will feed both the GMACS
medium-resolution optical spectrograph and the G-CLEF high-resolution
optical spectrograph (see paper by Jacoby et~al.\ in these proceedings).
The image-slicing capability means that MANIFEST more than doubles the
normal slit-limited spectral resolution of GMACS, providing a much
better match to the resolved velocity structure in the Lyman-$\alpha$
forest.

GMT+MANIFEST could be used to perform a large survey to construct a 3D
map of the IGM, using Lyman break galaxies (LBGs) as background sources
with which to probe the Lyman-$\alpha$ forest. The crucial wavelength
range for such a survey is 0.36--0.56\,$\mu$m, corresponding to a
redshift range $2.0<z<3.5$. The intrinsic size of the LBGs is
approximately 2\,kpc or 0.3\,arcsec, so they are well matched to the
fiber sampling of MANIFEST, with each starbug having 17 fibers each
sampling 0.2\,arcsec. With 4-hour exposures, GMT+MANIFEST can achieve
sufficient S/N for targets at $r \approx 25$, enabling a
$\sim$1\,deg$^2$ survey of LBG targets (with a surface density of
$\sim$4000 deg$^{-2}$ to $r \approx 24$) to be achieved in about 20
nights. This would be complemented by a sparsely-sampled galaxy redshift
survey of the same area, to a depth of $r \approx 26$ (corresponding to
a surface density of $\sim$30,000\,deg$^{-2}$), taking about 40 nights.
Such a combined survey would provide constraints on the growth of
small-scale structure at high redshifts that are not achievable with
8-metre class telescopes that have to use brighter (and sparser) QSOs as
the background sources for the Lyman-$\alpha$ forest.

In the ELT era, the greatest advantage of 8-metre-class telescopes will
be their relatively large fields of view. Thus, like 4-metre telescopes
in the 8-metre era, 8-metre telescopes in the ELT era will have strong
reasons to develop wide-field capabilities. Since LSST will dominate the
8-metre imaging niche, other 8-metre telescopes will likely seek to
develop some variety of wide-field MOS capability as their cutting-edge
instrumentation, as Subaru is already doing with PFS.

\section{CONCLUSIONS}

The contributions of multi-object spectroscopy to cosmology over the
past two decades have been profound. Redshift surveys of large-scale
structure are one of the pillars supporting modern precision cosmology,
and have considerable potential to increase their precision and range of
applicability. MOS instruments on 4-metre, 8-metre, and the next
generation of extremely large telescopes will continue to be powerful
tools for determining the evolution of the universe and attacking the
fundamental question of the nature of dark energy.

\acknowledgments      
 
This work supported in part by Australian Research Council grants
LE140100052 \& DP160102075. The author thanks his collaborators on the
2dFGRS, 6dFGS, WiggleZ, and Taipan science teams and on the 2dF, 6dF,
OzPoz, TAIPAN and MANIFEST instrument teams for making the last 20 years
both productive and pleasurable.

\bibliographystyle{spiebib} 
\bibliography{references}   

\end{document}